\documentclass[a4paper,pra,superscriptaddress,preprint,amsmath,amssymb,
 showpacs,nofootinbib]{revtex4}
\usepackage{graphicx}

\begin{document}
\begin{flushright}
{\small Ref. OU-HET 763/2012}
\end{flushright}

\title{Dynamics of two-photon paired superradiance}


\author{M.~Yoshimura}
\email{yoshim@fphy.hep.okayama-u.ac.jp}
\affiliation{Center of Quantum Universe, Faculty of Science, 
 Okayama University, Tsushima-naka 3-1-1 Kita-ku, Okayama 700-8530, Japan}
\author{N.~Sasao}
\email{sasao@fphy.hep.okayama-u.ac.jp}
\affiliation{Research Core for Extreme Quantum World, Okayama University,
 Tsushima-naka 3-1-1 Kita-ku, Okayama 700-8530, Japan}
\author{M.~Tanaka}
\email{tanaka@phys.sci.osaka-u.ac.jp}
\affiliation{Department of Physics, Graduate School of Science, 
 Osaka University, Toyonaka, Osaka 560-0043, Japan}

\begin{abstract}
We develop for dipole-forbidden transition
a dynamical theory of two-photon paired superradiance, or PSR for short.
This is a cooperative process 
characterized by two photons back to back
emitted with equal energies.
By irradiation of trigger laser from two target ends,
with its frequency 
tuned at the half energy between two levels,
a macroscopically coherent state of medium and fields dynamically emerges
as time evolves and large signal of amplified output
occurs with a time delay.
The basic semi-classical
equations in 1+1 spacetime dimensions are
derived for the field plus medium system to describe the spacetime
evolution of the entire system, 
and numerically solved to demonstrate existence of both explosive and weak PSR phenomena
in the presence of relaxation terms.
The explosive PSR event terminates accompanying  a sudden release of
most energy stored in the target.
Our numerical simulations are performed using a
vibrational transition $X^1\Sigma_g^+ v=1 \rightarrow 0$ of para-H$_2$  molecule,
and taking many different excited atom number densities
and different initial coherences between the metastable and the ground states.
In an example of 
number density close to $O[10^{21}]$cm$^{-3}$ 
and of high initial coherence,
the explosive event terminates at several nano seconds after the trigger
irradiation, when the phase relaxation time of $> O[10]$ ns is taken.
After PSR events the system is expected to follow a steady state solution
which is obtained by analytic means, and 
is made of many objects of
field condensates endowed with a topological stability.

\end{abstract}


\maketitle

\section{Introduction}

\begin{figure*}[htbp]
\includegraphics[width=20em]{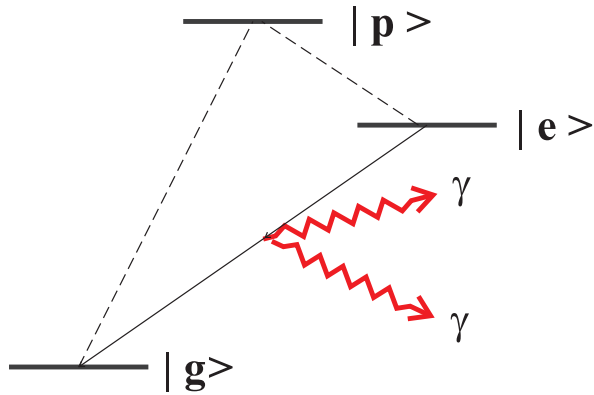} 
\caption{\label{lambda-type atom}
(Color online)
 $\Lambda-$type  atomic level for PSR. Dipole forbidden transition 
 $|e \rangle \rightarrow |g\rangle + \gamma + \gamma$ 
 may occur via strong E1 couplings to $|p\rangle$.}
\end{figure*}

Since the early suggestion \cite{first 2g}
a variety of coherent two-photon processes have attracted
much interest, both from theoretical 
\cite{ladder 2g}, \cite{narducci}, \cite{harris} and experimental sides
\cite{ladder 2g-2}, \cite{ladder 2g-3}, \cite{ladder 2g-4}, \cite{eit review}, \cite{2g laser}.
Our present work is focused on a different aspect 
of coherent two photon emission from
$\Lambda-$type three level atoms (or molecules) where
transition between
two lower levels is dipole-forbidden (see Fig(\ref{lambda-type atom}) for 
the level structure).
As pointed out in \cite{macro-coherence},
a macroscopic target made of
metastable atoms in the $|e\rangle$ of Fig(\ref{lambda-type atom})
may induce a characteristic event of macro-coherent two photon emission,
two photons exactly back to back emitted with equal energies.
We use for this phenomenon the terminology of two-photon paired superradiance, or PSR in short.
The term paired is used
because two emitted photons are highly correlated in their momenta and
spin orientations (most clearly seen in $J=0 \rightarrow 0$ transition).
The rate enhancement factor in the momentum configuration of the back to back
emission is expected much larger
than in the usual superradiance (SR) case \cite{sr review}
due to lack of the wavelength limitation there:
the coherent volume for SR
is limited with the wavelength $\lambda$ by $\lambda^2 L$ where $L$ is the target
length for a cylindrical configuration, while the macro-coherent PSR
has the coherent volume of entire cylinder irradiated by trigger.

The usual single photon superradiance occurs irrespective of
absence or presence of trigger  due to the intrinsic instability
of exponential spontaneous decay caused by dipole-allowed transition.
On the other hand,  two-photon emission
occurs with a much smaller rate in higher order of perturbation
beyond dipole-forbidden transition,
hence the use of trigger is essential to assist 
the macro-coherence development for two-photon process
and induce rapid PSR events of large signal.
Quantum initiation such as proposed for SR in \cite{quantum initiation}
is not needed, since PSR is more akin to the triggered SR \cite{triggered sr},
which makes appropriate the following semi-classical treatment.

A basic formalism of two-photon process already exists,
\cite{narducci} for propagation equation and \cite{my-10-10}
for analytic results of the propagation problem 
and PSR emission treated as perturbation.
But this formulation turns out insufficient to
dynamically discuss two (back to back) mode propagation
incorporating PSR, 
which seems essential for dense medium.
In the present work
we shall be able to derive a fundamental set
of semi-classical equations for the two mode and further present
formulation of two color problem as well.
The essential ingredient in our work
is derivation of a more general quantum mechanical equation both
for the medium (Bloch equation) and the
electromagnetic field (extension of quantum Maxwell
equation to include two photon process).
Only after elucidating the nature of quantum state of fields and medium,
namely, time evolving electric field condensate,
we shall go on to the semi-classical equation.
This way we determine how two back to back modes are precisely coupled
beyond the perturbation theory.

We ignore the granularity and rely on the continuum limit formulation of
atom distribution, taking
one spatial dimension alone, because
the whole event is highly focused on one direction 
of irradiated trigger field taken as $x$ direction.
The system of semi-classical partial
differential equations thus derived is highly non-linear, 
and must generally be analyzed by numerical simulations.
This way we find explosive and weak PSR phenomena and
under what conditions these may occur.

Despite of its complicated non-linearity
the system allows soliton solution of
two kinds, which
is obtained as steady state solutions of this non-linear system
of fields and medium.
Solitons here, in their field part, are electric field condensate which may or may
not be moving: there can be static field condensate.
The stability (against two-photon emission) of solitons is ensured by a topological
quantum number as explained in the text below.
Our conjecture, which is supported by numerical simulations,
but not established by a more rigorous method,
is that field condensate formed after rapid PSR phenomena is made of 
many topological solitons.
After formation of field condensates,
namely a stable target state against two photon emission,
the light may  propagate almost freely.
The condensate state of field plus medium thus formed may be very useful 
to detect a much weaker process such as radiative
neutrino pair emission (RNPE) \cite{my-rnpe},
because the condensate is not stable against RNPE.

A related propagation and soliton formation problem in the single photon case
is the phenomenon of
self-induced transparency (SIT) \cite{coherent light propagation in 2 level}
and electromagnetically induced transparency (EIT) \cite{eit review}
presumably related to solitons of the kind of \cite{yavuz},
both at a resonant frequency.
Both of these transparency phenomena thus appear to be directly related to
formation of stable solitons of different kinds from ours.

For numerical computations below, we use parameters
relevant to a good target candidate for PSR detection,
para-H$_2$ molecule.
We have in mind using para-H$_2$ vibrational transition of 
$X^1\Sigma_g^+ v=1 \rightarrow 0$ (X being the electronically
ground molecular state).
Many other atoms and molecules are conceivable
for PSR experiments.
The characteristic length scale for large effects is $\sim 14 $cm, and
the time scale $\sim 0.5$ ns for para-H$_2$ 
of a molecule density of $n=10^{20}$cm$^{-3}$.
The number density dependence of these characteristic
parameters is  $\propto 1/n$.
We include relaxation effects of two time constants in the range of
$T_2 \geq$ 10 ns (a feasible value experimentally)
and $T_1\gg T_2$ in our analysis.
Origin of these relaxation constants is left unexplained, and
this way one may use values experimentally measured  by
other means.
We perform extensive numerical simulation
in order to clarify experimentally observable PSR signals and condensate
formation in forthcoming experiments.
It is demonstrated that explosive PSR emission occurs
for long targets even by weak trigger  when initial coherence between
states, $|e\rangle$ and $|g\rangle$, is present. 
We have identified two different types of PSR events caused by trigger irradiation:
(1) explosive PSR in which most of the stored energy in the
initial metastable state $|e\rangle$
is released as a short pulse of some time structure, and
(2) weak PSR in which the output energy flux is in linear proportion to
the trigger power.

The natural unit $\hbar = c= 1$ is used 
throughout in the present paper.

\section{Derivation of quantum and semi-classical equation}

Consider three level atom (or molecule) of energies, 
$\epsilon_p > \epsilon_e >  \epsilon_g$, as shown in
Fig(\ref{lambda-type atom}).
We assume that transition between two lower levels, $|e\rangle$ and $|g\rangle$,
are dipole forbidden.
Suppose that the upper level $|p\rangle$ has
substantial E1 rates both to  $|e\rangle\,, |g\rangle$.
(This can be replaced by weaker M1 transition, 
since  the relation we need subsequently is 
the partial decay rate $ \propto \epsilon_{ij}^3$ 
with the energy level difference $\epsilon_{ij}=\epsilon_i - \epsilon_j$,
which holds both in E1 and M1 cases.)

We focus on,  and derive 
an effective hamiltonian of, two lower levels
interacting with oscillating electric field $E$.
Its hamiltonian
density has been derived in \cite{narducci}, \cite{my-10-10}
for a single mode of field such as a light wave of definite frequency
traveling in one direction.
Extension to multi-mode fields such as counter-propagating modes
of the same frequency is given in Appendix A.
Its hamiltonian has a form of $2\times 2$ matrix
acting on two atomic states, $|e \rangle$ and $|g \rangle$,
\(\:
\sim E^T {\cal M} E
\:\).
The multi-mode field $E$ may be decomposed into
positive and negative frequency parts
\(\:
E^T = \sum_j \frac{1}{2}(E_j^* e^{i\omega_j t} + E_j e^{-i\omega_j t})
\:\),
where $E_j, E_j^*$ are slowly varying envelopes in time.
We shall use variables,
\(\:
E_j^{+} = E_j e^{-i\omega_j t}
\,, \hspace{0.2cm}
E_j^{-} = E_j^* e^{i\omega_j t}
\,,
\:\)
to simplify formulas given below.
In quantum field theory $E_j$ and $E_j^*$
represent annihilation and creation operators of definite mode.
The pertinent hamiltonian to our discussion of the single mode is
\begin{eqnarray}
&&
\frac{d}{dt}
\left(
\begin{array}{c}
c_e (x,t) \\
c_g  (x,t)
\end{array}
\right)
= - i{\cal H}_I \left(
\begin{array}{c}
c_e (x,t) \\
c_g  (x,t)
\end{array}
\right)
\,,
\label{atom-field interaction}
\\ &&
- {\cal H}_I =
\left(
\begin{array}{cc}
 \mu_{ee} E^{+}E^{-}
& e^{i \epsilon_{eg}t} \mu_{ge} (E^{+})^2  \\
e^{-i \epsilon_{eg} t} 
\mu_{ge} (E^{-})^2
&   \mu_{gg} E^{+}E^{-}
\end{array}
\right)
\,,
\\ &&
\mu_{ge} = \frac{2d_{pe}d_{pg}}
{\epsilon_{pg}+\epsilon_{pe}}
\,, \hspace{0.5cm}
\mu_{aa}= \frac{2d_{pa}^2\epsilon_{pa}}{\epsilon_{pa}^2 - \omega_0^2}
\,, \hspace{0.5cm} (a = g\,, e)
\label{1 mode mu}
\,,
\end{eqnarray}
where $|c_e|^2 + |c_g|^2=n(x)$ with
$n(x)$ the number density of atoms per a unit volume 
in a linear target region of $0\leq x \leq L$.
For simplicity we took isotropic medium and
linearly polarized fields, taking $\vec{E}^{\pm}$ as scalar
functions.
The diagonal part $\propto \mu_{aa}$ of this hamiltonian describes
AC Stark energy shifts, while off-diagonal parts $\propto \mu_{ge}$ are
for two photon emission and absorption.

For pH$_2$ target the photon energy $\omega_0=\epsilon_{eg}/2\sim$ 0.26 eV
is much smaller than level spacings to the electronically excited
intermediate states,  both $\epsilon_{pe}$ and $ \epsilon_{pg}\sim $ 11 eV.
Under this condition we may ignore $\omega_0$ compared to
$\epsilon_{pa}\,, a=e, g$ in the formula for $\mu_{ab}$ and
identify $\mu_{ab}$ to the polarizability for which
precision calculation exists \cite{ph2 polarizability}.
We thus use numerical values of parameters, 
\(\:
\mu_{gg}\sim 0.80\,,
\mu_{ee} \sim 0.87\,,
\mu_{ge} \sim 0.055
\:\)
all in the unit of
\(\:
10^{-24}
\:\)
cm$^{3}$ \cite{atomic parameter}
for the pH$_2\; Xv=1 \rightarrow Xv=0$ transition.

The density matrix 
of pure atomic
states, $\rho = | \psi(x,t)\rangle \langle \psi(x,t)|$
($\langle \psi(x,t)| = (c_e,c_g)$), 
obeys the evolution equation,
\(\:
\partial_t \rho = -i[{\cal H}_I,\rho]
\,.
\:\)
This quantum mechanical equation is generalized
to include dissipation or relaxation.
The needed variable, the density matrix for the mixed state,
is given by a statistical mixture of pure states:
\begin{eqnarray}
&&
\rho (x,t) = \sum_i c_i | \psi_i(x,t)\rangle \langle \psi_i(x,t)|
\,,
\hspace{0.5cm}
\sum_i c_i = 1
\,, \hspace{0.5cm}
0\leq c_i \leq 1
\,,
\end{eqnarray}
with $| \psi_i(x,t)\rangle$ a set of orthonormal pure state vectors.
Dissipation occurs when a subsystem of $|e \rangle \,, |g \rangle$ interacts
with a reservoir and one integrates out reservoir
variables due to our basic ignorance of the reservoir.
The general form of mixed state evolution
including dissipation has been
derived by Lindblad \cite{lindblad}, 
assuming the general principle of positivity and conservation
of probability.
As its result the time evolution equation of the density matrix 
has additional operator term, $L[\rho]$.
The new additional dissipation term in the two level atomic
system turns out equivalent to phenomenological relaxation terms
given by two time constants, $T_1, T_2$ (with 
the constraint $T_1 > T_2/2$ from consistency with \cite{lindblad}).

It is convenient to write the evolution equations in terms
of components of the Bloch vector defined by
$\vec{R}={\rm tr}\; \rho \vec{\sigma}=\langle \psi|\vec{\sigma}|\psi \rangle$.
The basic Bloch equation including relaxation terms is 
\begin{eqnarray}
\!\!\!\partial_t R_1 &\!=\!&
 (\mu_{ee}-\mu_{gg})E^+E^- R_2 - i\mu_{ge} 
(e^{i\epsilon_{eg}t}E^+E^+ - e^{-i\epsilon_{eg}t}E^-E^-) R_3
-\frac{R_1}{T_2}
\,,
\label{bloch eq 1} 
\\
\!\!\!\partial_t R_2 &\!=\!& -(\mu_{ee}-\mu_{gg})E^+E^- R_1
+ \mu_{ge} (e^{i\epsilon_{eg}t}E^+E^+ + e^{-i\epsilon_{eg}t}E^-E^-) R_3 
-\frac{R_2}{T_2}
\,,
\label{bloch eq 2}
\\
\!\!\!\partial_t R_3 &\!=\!&
 \mu_{ge} \left(
i  (e^{i\epsilon_{eg}t}E^+E^+ \!-\! e^{-i\epsilon_{eg}t}E^-E^-  ) R_1
\!-\!  (e^{i\epsilon_{eg}t}E^+E^+ \!+\! e^{-i\epsilon_{eg}t}E^-E^-  )R_2
\right) 
\!-\!\frac{R_3 + n}{T_1}.
\label{bloch eq 3}
\end{eqnarray}
$T_1 \gg T_2$ usually, and the phase decoherence time $T_2$
is much smaller and more important than the decay time $T_1$,
which may be taken infinitely large for our practical purpose.

Derivation of quantum field equation follows a similar line of reasonings.
To perform the derivative operation $\partial_t^2$ as in the Maxwell equation,
one needs to calculate the double commutator;
\begin{eqnarray}
&&
\partial_t^2 \vec{E}^{\pm} =
- [H \,, [H \,,  \vec{E}^{\pm} \,]\,]
\,, \hspace{0.5cm}
H = \int d^3x ({\cal H}_f + {\rm tr}\, \rho {\cal H}_I) \,,
\end{eqnarray}
with the field energy density ${\cal H}_f = (\vec{E}^2+\vec{B}^2)/2$.
For convenience we add less dominant oscillating terms of field modes to
$E^{\pm}$ and use the locally well behaved field $E(x,t)$ in ${\cal H}_I$.
The fundamental commutation relation in the radiation gauge QED
$[E_y (\vec{r}, t)\,, B_z(\vec{r}', t)] 
= i \partial_x \delta^3 ( \vec{r}- \vec{r}\,') $ 
\cite{jj advanced} is used
for derivation of quantum field equation.
The result is
\begin{eqnarray}
&&
\hspace*{1cm}
(\partial_t^2 - \vec{\nabla}^2 )\vec{E}^{\pm}
=   \vec{\nabla}^2 {\cal D} \vec{E}^{\pm}
\,,
\label{full field eq}
\\ &&
\hspace*{-1cm}
-{\cal D}\vec{E}^+
=
\left( \frac{\mu_{ee} + \mu_{gg}}{2}
n + \frac{\mu_{ee} - \mu_{gg}}{2}
R_3  \right)\vec{E}^+ + \mu_{ge} e^{-i \epsilon_{eg}t}
(R_1-iR_2) \vec{E}^-
\,.
\label{rhs of field eq}
\end{eqnarray}
This equation \cite{another derivation} along with the Bloch
equations (\ref{bloch eq 1})$\sim$ (\ref{bloch eq 3}) 
is the basis of the following derivation of our master equation.

\vspace{0.5cm}
{\bf Slowly Varying Envelope Approximation (SVEA)}
\hspace{0.2cm}
Fast oscillating terms do not contribute to
global features of time and spatial evolution
when one makes averaging over a few $\times$ time and spatial 
oscillation periods.
We thus extract terms that persist over time periods of
typical light oscillation of order $1/\omega$ both in time and
space.
Envelope functions denoted by $E_R\,, E_L$ ought to be amplitudes of right and
left moving components of rapidly oscillating
parts $\propto e^{-i\omega (t \mp x)}$.

The result of SVEA may be summarized using dimensionless units
of spacetime coordinates $\xi\,, \tau$ and 
dimensionless fields $e_{L, R}$
given by
\begin{eqnarray}
&&
(\xi\,, \tau) =  (\alpha_m x\,, \alpha_m t)
\,,\hspace{0.5cm}
\alpha_m(\omega) = \frac{\epsilon_{eg}}{2}n\mu_{ge}(\omega)  
\,, \hspace{0.5cm}
|e_{L, R}|^2 = \frac{|E_{L, R}|^2}{\epsilon_{eg}n}
\,, \hspace{0.5cm}
r_i = \frac{R_i}{n}
\,.
\end{eqnarray}
The quantity $1/\alpha_m = 2/(\mu_{ge}\epsilon_{eg}n)$ 
gives a fundamental unit of target length and time scale of evolution.
Since a functional relation $\alpha_m(\omega)= \alpha_m(\epsilon_{eg}-\omega)$
holds,
the propagation problem of trigger irradiation of pair frequencies, 
$\omega$ and 
$\epsilon_{eg}-\omega$, is described by the same dimensionless quantities
of a common $\alpha_m$.
Its value at $\omega=\epsilon_{eg}/2$ is $\sim 14 $cm and $\sim$ 0.5 ns for 
para-H$_2$ 
of density $ 10^{20}$cm$^{-3}$.

The most general fundamental equations including
both non-trivial propagation and PSR effects are derived
in Appendix A and given by the formulas (\ref{rescaled bloch eq1}) $\sim$ 
(\ref{rescaled quantum field eq2}).
It is useful to recall the physical meaning of coupling constants $\mu_{ab}$
in the interaction hamiltonian, in order to fully appreciate the following
approximation in our numerical simulations.
Consider the extended hamiltonian including both of counter-propagating modes
given by eq.(\ref{1c two modes hamiltonian}) in Appendix. 
We first note that annihilation ($a_i$) and creation ($a_i^{\dagger}$) operators 
of photon modes are related to complex fields by
$E_i^+ \sim a_i \sqrt{\omega/2V}\,, E_i^- \sim a_i^{\dagger} \sqrt{\omega/2V}$
where $V$ is the quantization volume.
The important equations are obtained after SVEA and given in Appendix A.
They are written in terms of envelope functions:
\begin{eqnarray}
&& 
(\partial_t + \partial_x) E_R =
\frac{i\omega}{2}\biggl( (\frac{\mu_{ee} + \mu_{gg}}{2}
n + \frac{\mu_{ee} - \mu_{gg}}{2}R_3^{(0)})E_R 
+ \frac{\mu_{ee} - \mu_{gg}}{2}R_3^{(+)}E_L 
\nonumber \\ &&
\hspace*{1cm}
+ \mu_{ge} 
\left( (R_1-iR_2)^{(0)} E_L^* + (R_1-iR_2)^{(+)}E_R^* \right)
\biggr)
\,,
\label{lr mode coupling 1}
\\ &&
(\partial_t - \partial_x) E_L =
\frac{i\omega}{2}\biggl( (\frac{\mu_{ee} + \mu_{gg}}{2}
n + \frac{\mu_{ee} - \mu_{gg}}{2}R_3^{(0)})E_L 
+ \frac{\mu_{ee} - \mu_{gg}}{2}R_3^{(-)}E_R 
\nonumber \\ &&
\hspace*{1cm}
+ \mu_{ge} 
\left( (R_1-iR_2)^{(0)} E_R^* + (R_1-iR_2)^{(-)}E_L^* \right)
\biggr)
\,.
\label{lr mode coupling 2}
\end{eqnarray}
The right hand sides of  these equations give effects, all in bulk medium, 
of forward scattering  $\propto \frac{\mu_{ee} + \mu_{gg}}{2}
n + \frac{\mu_{ee} - \mu_{gg}}{2}R_3^{(0)}$,
backward scattering  $\propto \frac{\mu_{ee} - \mu_{gg}}{2}R_3^{(\pm)}$,  
RL- pair annihilation $\propto \mu_{ge}  (R_1-iR_2)^{(0)} $,
and RR-, LL-pair annihilation  $\propto \mu_{ge} (R_1-iR_2)^{(\pm)}$.
(The pair creation amplitudes appear in conjugate equations to those above.)
Quantities $R_i ^{(\pm)}e^{\pm 2i kx}$ as defined by eq.(\ref{case of spatial graiting})
are what are called spatial grating in the literature.
The backward scattering terms, and RR,- LL-pair annihilation
and creation terms  are important only in the presence of
spatial grating of polarization.
Neglect of spatial grating is thus equivalent to retaining 
forward scattering  and RL-pair processes, and ignoring
all other terms.
In the simple boundary condition set up below in this work,
the backward Bragg scattering  is expected to be a minor effect, and
is also neglected in most works of SR.
We refer to \cite{lewenstein} on the backward Bragg scattering effect in usual SR,
and for instance to \cite{bragg in mode sr} on the backward scattering effect
on SR in low-Q cavity experiments. 
In a more comprehensive simulation in future we wish to quantitatively compute
effects of the backward scattering, RR-, LL-pair processes,
because non-negligible differences of these effects might arise in PSR unlike the SR case.

In the rest of the present work we shall focus on
PSR effects and ignore propagation effects which are much
discussed in \cite{narducci}, \cite{my-10-10}
and summarized in Appendix A.
Explosive PSR events discussed below are
expected to be insensitive to neglected propagation effects.
The resulting Maxwell-Bloch equation for the single mode is 
\begin{eqnarray}
&&
\partial_{\tau} r_1 = 4\gamma_-(|e_R|^2+|e_L|^2)r_2 
+ 8 \Im (e_R e_L)r_3
-\frac{r_1}{\tau_2}
\,,
\label{rescaled bloch eq degenerate 1}
\\ &&
\partial_{\tau} r_2 = 
-4\gamma_-(|e_R|^2+|e_L|^2)r_1
+8 \Re (e_R e_L )r_3
-\frac{r_2}{\tau_2}
\,,
\\ &&
\partial_{\tau} r_3 = 
-8 \left( \Re (e_R e_L)r_2  +
\Im (e_R e_L )r_1
\right)
-\frac{r_3+1}{\tau_1}
\,,
\label{rescaled bloch eq degenerate 3}
\end{eqnarray}
\begin{eqnarray}
&&
(\partial_{\tau} + \partial_{\xi})e_R = 
 \frac{i}{2}  (\gamma_+  +  \gamma_- r_3 ) e_R
+ \frac{i}{2}(r_1 - ir_2)e_L^*
\,, 
\label{rescaled quantum field eq: degenerate 1}
\\ &&
(\partial_{\tau} - \partial_{\xi})e_L = 
 \frac{i}{2} (\gamma_+  +  \gamma_- r_3 ) e_L
+ \frac{i}{2}(r_1 - ir_2) e_R^* 
\,,  
\label{rescaled quantum field eq: degenerate 2}
\end{eqnarray}
\begin{equation}
\gamma_{\pm} = \frac{\mu_{ee} \pm \mu_{gg}}{2\mu_{ge}}
\,.
\end{equation}
Here $\tau_i = \alpha_m T_i$ are relaxation times in
the dimensionless unit.

The dimensionless master equation (\ref{rescaled bloch eq degenerate 1}) 
$\sim$ (\ref{rescaled quantum field eq: degenerate 2})
is governed by two important parameters, the most important is 
$\tau_2= \alpha_m T_2 $
and the next important is $\gamma_{\pm}$.
Another experimentally important parameter is the overall length and
time $1/\alpha_m \propto 1/n$, inversely scaling with the number density $n$.
For larger number densities of excited atoms
a smaller size target and a shorter time measurement of O[ns]
becomes possible.

In terms of two component field $\varphi$ defined below the equation reads as
\begin{eqnarray}
&&
(\partial_{\tau} + \sigma_3\partial_{\xi})\varphi 
= \frac{i}{2} (\gamma_+ + \gamma_- r_3) \varphi
+ \frac{i}{2}(r_1 - ir_2)  \sigma_1 \varphi^{\,*}
\,, \hspace{0.5cm}
\varphi = 
\left(
\begin{array}{c}
 e_R \\
  e_L
\end{array}
\right)
\,.
\label{quantum field eq 1}
\end{eqnarray}
Magnitudes of R- and L-fluxes change via RL mixing term such as
\begin{eqnarray}
&&
(\partial_{\tau} \pm \partial_{\xi})|e_{R\,, L}|^2 = 
r_1 \Im (e_R e_L) +r_2 \Re (e_R e_L)
\,.
\end{eqnarray}
R- or L-moving pulse alone propagates freely, 
because we ignored in this approximation
non-trivial propagation effects.

\vspace{0.5cm}
{\bf  Quantum state of fields}
\hspace{0.2cm}
As usual in quantum field theory, 
we may interpret $\vec{E}_{R,L}$ as annihilation
and $\vec{E}_{R,L}^{\,\dagger}$ as creation operator.
The fact that the basic equation, 
(\ref{rescaled quantum field eq: degenerate 1}) $\sim$
(\ref{rescaled quantum field eq: degenerate 2}) or (\ref{quantum field eq 1}),
simultaneously contains both
annihilation and creation operators of field implies
that the quantum state satisfying the field equation
is given by a Bogoliubov transformation from the usual vacuum 
of zero photon state $|0\rangle$,
\begin{eqnarray}
&&
| \Psi \rangle = 
\sum_{n=0}^{\infty} c_n(x,t) (E_R^{\,\dagger}E_L^{\,\dagger})^n |0\rangle
\,,
\label{bogoliubov state}
\end{eqnarray}
where
$c_n(x,t)$ is to be determined by 
\(\:
\left( {\rm eq.}(\ref{quantum field eq 1}) \right)
|\Psi\rangle = 0.
\:\)
The quantum state $|\Psi\rangle$ is a mixture of
infinitely many states of different photon number.
We shall not pursue this line of thoughts any further,
because we exploit the semi-classical approximation
under the large quantum number limit of photons (the classical limit).
The semi-classical equation is given by the
expectation value of quantum equation,
\begin{eqnarray}
&&
\langle \Psi | \left(
(\partial_{\tau} + \sigma_3\partial_{\xi})\varphi 
-\frac{i}{2} (\gamma_+ + \gamma_- r_3) \varphi
- \frac{i}{2}(r_1- ir_2)  \sigma_1 \varphi^{\,*}
\right)
|\Psi \rangle
=0 \,,
\end{eqnarray}
with $|\Psi \rangle$ the Bogoliubov state given by eq.(\ref{bogoliubov state}).
The semi-classical equation turns out equivalent to
replacing q-field operators in the quantum equation by corresponding 
c-number functions.
Equations, (\ref{rescaled bloch eq degenerate 1}) $\sim$ 
(\ref{rescaled quantum field eq: degenerate 2}), regarded
as equations for  c-number functions, thus
constitute the master equation
for polarization of medium and field.

In our case of field condensate,
 medium polarization and fields are cooperatively involved:
the target medium undergoes coherence oscillation,
simultaneous with field oscillation, while keeping 
field envelopes slowly varying and finally almost time independent in a large
time limit, as shown below.
The field condensate part is technically equivalent to
field state   made of an infinite sum
of multiple photon pair states in the so-called coherent state representation.

\section{Importance of initial coherence}
It is important to clarify in detail the ideal case of numerical solutions
where all quantities in eq.(\ref{rescaled bloch eq degenerate 1}) 
$\sim $ (\ref{rescaled quantum field eq: degenerate 2})
are of order unity, in the range of $O[10 \sim 1/10] $. 
For a deeper understanding of numerical outputs and
a practical check of accuracy of numerical results,
it is useful to know conservation laws of our non-linear system.
We list in Appendix B all exact and approximate conservation
laws that the system possesses.

We have performed numerical simulations assuming CW
(non-pulsed continuous wave)
trigger laser irradiation of the same power from two target
ends (called the symmetric trigger).
This boundary condition is similar, but not identical,
to the one of cavity mirror.
Use of cavity mirrors has both advantage and complication.
Two mirrors in cavity automatically generate counter-propagating
waves, and they effectively increase the trigger power
(which however is not critically needed in our case).
On the other hand, each atom in cavity is affected
by the same traveling fields many times and this
complicates analysis.
We use in the present work the simpler scheme of two CW counter-propagating 
triggers independently irradiated.

\begin{figure*}[htbp]
\includegraphics[width=30em]{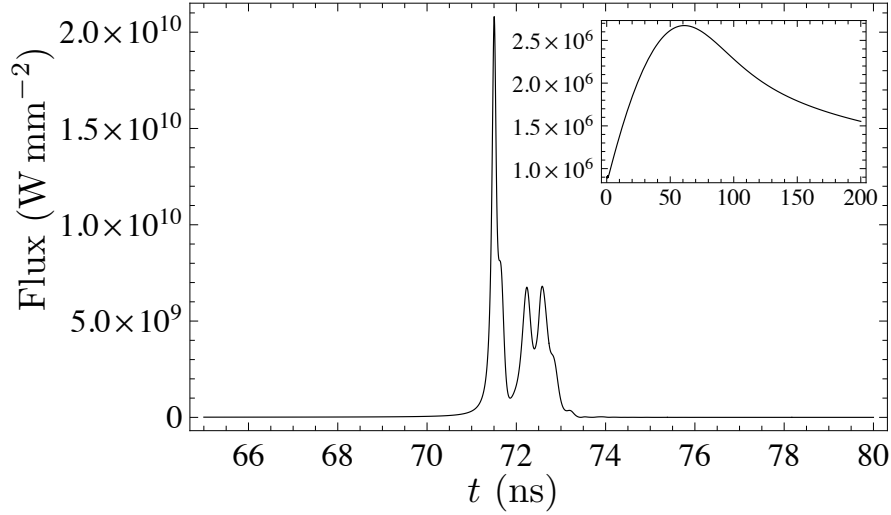}
\caption{\label{CW strong trigger} 
 Time evolving output flux at a target end of length 30 cm
 resulting from the symmetric CW trigger irradiation of power 
 1 ${\rm MW\,mm}^{-2}$ and 0.9 ${\rm MW\,mm}^{-2}$ in the inset.
 (Note a large difference of $\sim 10^4$ of the output power in two plots.)
 Assumed parameters are $n=1\times 10^{21} {\rm cm}^{-3}$ of pH$_2$,
 numerical values (see the text) of $\mu_{ab}$ for the transition 
 $Xv=1 \rightarrow Xv=0$, relaxation times $T_2=10,\; T_1=10^3$ ns's, 
 and initial complete inversion (hence no coherence) of $r_3=1$ with 
 $r_1,\,r_2=0$ taken for the initial target state.}
\end{figure*}

Numerical results show the symmetric output fluxes from two ends,
and we exhibit in the following figures one of these
identical fluxes from one end.
Result for the zero initial coherence $r_1(\xi, 0) = r_2(\xi, 0) =0$ is
shown in Fig(\ref{CW strong trigger}).
Clear signature of delay much after $T_2$
($\sim 7 T_2$ in this case) and explosive PSR is observed
for strong trigger fields.
It is difficult to obtain commercially available CW laser
of this power.
A reason of this difficulty is that relaxation 
of order $T_2 \sim 10$ns may
take over the coherence development under weak trigger
usually exploited.
Explosive PSR is a highly non-linear process having a definite 
trigger power threshold
and disappears in this example certainly at the trigger power of
0.9 MWmm$^{-2}$, as shown in the inset of
Fig(\ref{CW strong trigger}).

In Fig(\ref{CW strong trigger}) the complete inversion to the level $|e\rangle$
has been assumed as an initial condition, 
and it would be interesting to relax this condition
and to further clarify neglected effects
of the presence of initial coherence between two atomic
levels, $|e\rangle$ and $|g\rangle$.
There is an experimental method to imprint an initial coherence
between $|e\rangle $ and $|g\rangle $
by adopting a clever excitation scheme.
The atomic state right after excitation
can be made a coherent mixture of two pure states,
$|e\rangle$ and $|g\rangle$, namely
$c_e|e\rangle + c_g|g\rangle $ with $ |c_e|^2+ |c_g|^2=1$
at a single atomic site,  by 
using the technique of STIRAP \cite{stirap}.
This kind of pure state may be formed by 
time overlapping excitation pulses of two frequencies,
$\approx \epsilon_{pe}$ and $\approx \epsilon_{pg}$.
The state is a dark state,
called so because no emission from $|p\rangle$ is observed
despite of irradiation capable of making both transitions,
$|p\rangle \rightarrow |e\rangle$ and $|p\rangle \rightarrow |g\rangle$.

The medium polarization $r_i$ in the dark state is given by
\begin{eqnarray}
&&
r_1 = 2\sqrt{p(1-p)}\cos \theta_0 
\,, \hspace{0.5cm}
r_2 = 2\sqrt{p(1-p)}\sin \theta_0
\,, \hspace{0.5cm}
r_3 = 2p-1
\,,
\label{dark state of ri}
\end{eqnarray}
with $p$ the fraction in the state $|e \rangle$.
When this type of initial polarization of the dark state is formed,
one may expect to expedite 
the coherence development for PSR, as shown in the following section.
When CW laser is used for trigger,
two overlapping pulses may induce PSR at the same time when
the emission from $|p\rangle $
disappears:  thus it may be called PSR from the dark.

\section{Numerical solutions for high density target with initial coherence}
We first comment on what the number density $n$ of target precisely means.
This is the total number of atoms/molecules per a unit volume
participating in PSR phenomena, hence 
it is the added sum of densities in the states,
$|e \rangle$ and $|g \rangle$.
Note also that 
the state $|g \rangle$ may or may not be the ground state of
atoms or molecules. For instance, in the pH$_2$ transition of 
$X^1\Sigma_g^+ v=2 \rightarrow 1$,
the target number density $n$ may be much less than the
ground state number density since $|g\rangle = (Xv=1)$ is also an
excited state.

\begin{figure*}[htbp]
\includegraphics[width=30em]{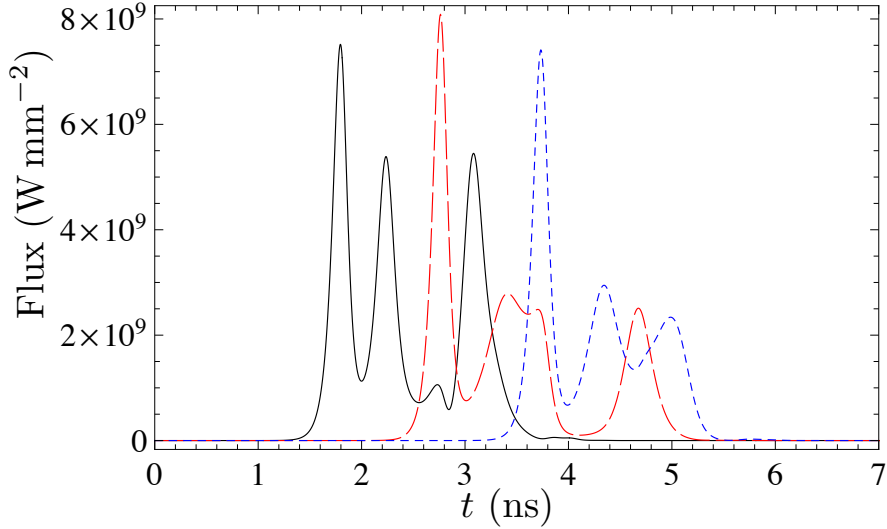}
\caption{\label{power dependence of PSR}
(Color online)
 Trigger power dependence of time-evolving output flux from the symmetric 
 trigger irradiation of the power range, $10^{-12}\sim 1$Wmm$^{-2}$, 
 under the conditions of $n=1\times 10^{21}$cm$^{-3}$, target length $= 30 $cm, 
 relaxation times $T_2=10, T_1=10^3$ ns's, and the initial polarization, 
 $r_1=1, r_2=r_3=0$. Depicted outputs from 1 Wmm$^{-2}$ trigger power in solid
 black, from $10^{-6}$Wmm$^{-2}$ in dashed red, and from $10^{-12}$Wmm$^{-2}$
 in dotted blue are displaced almost equi-distantly in the first peak positions.
 Transition $Xv=1 \rightarrow Xv=0$ of pH$_2$ is considered. $\sim $ 70 \% 
 stored energy in the initial metastable state is released in these cases.}
\end{figure*}

\begin{figure}[htbp]
\begin{minipage}{20em}
\includegraphics[width=\textwidth]{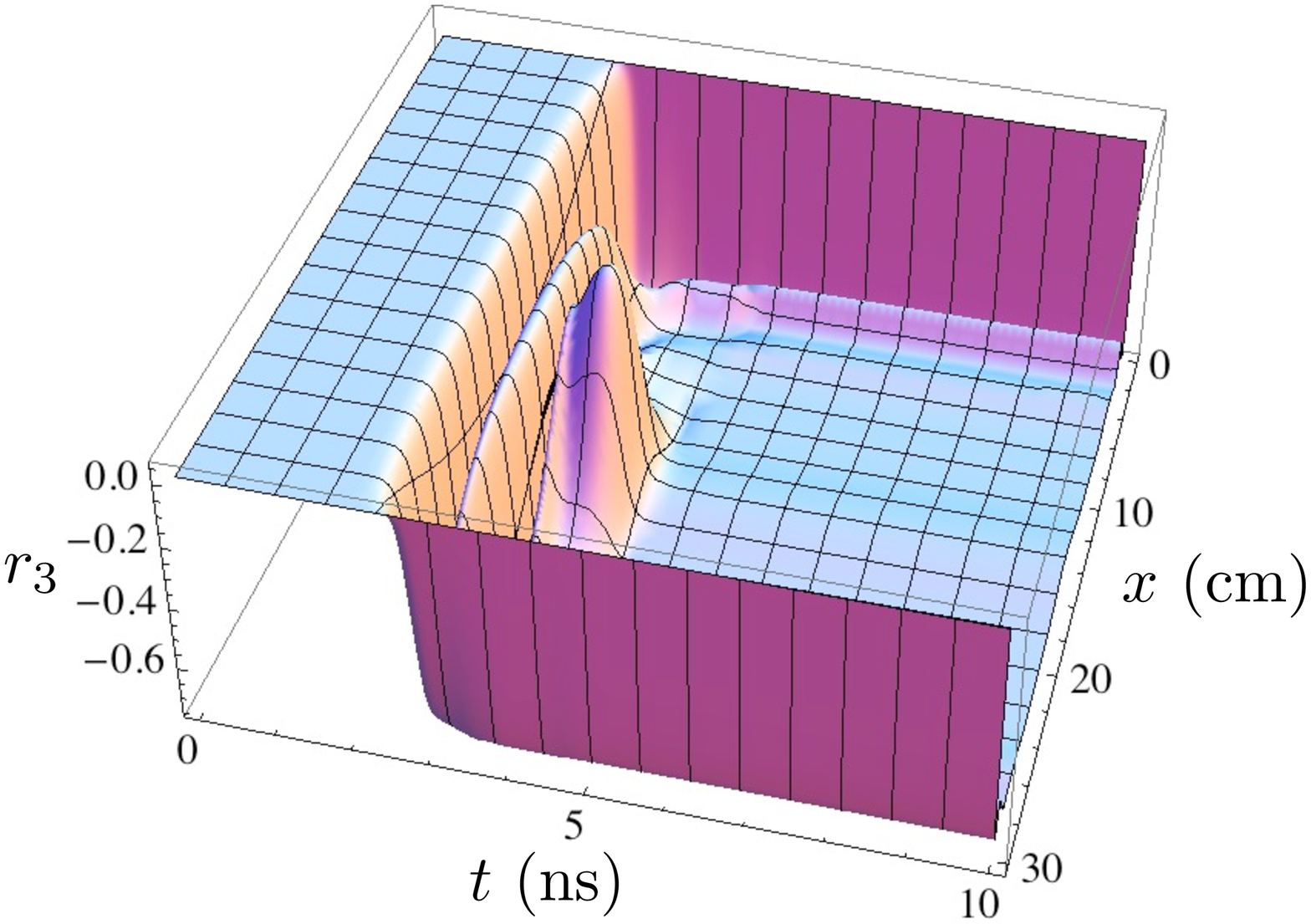}
\caption{\label{r3 profile}
(Color)
 Spacetime profile of $r_3$  for the 1 $\mu$Wmm$^{-2}$ case 
 of Fig(\ref{power dependence of PSR}).}
\end{minipage}
\hspace{2em}
\begin{minipage}{20em}
\includegraphics[width=\textwidth]{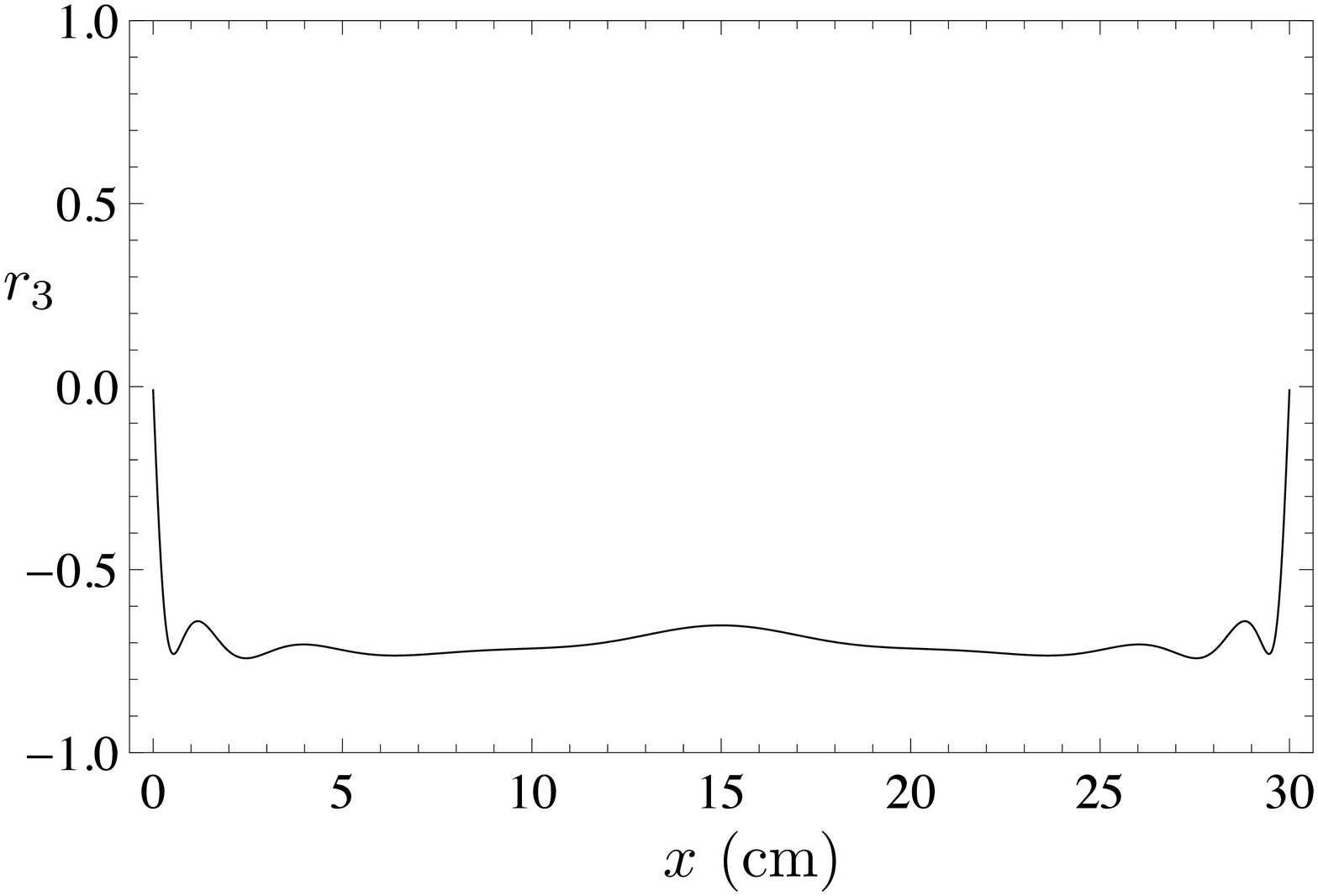}
\caption{\label{r3 profile 2}
 Spatial profile of $r_3$ at the latest time, 10 ns 
 after trigger irradiation, of Fig(\ref{r3 profile}).}
\end{minipage}
\end{figure}

Time evolution from a dark state of initial polarization value given by
eq.(\ref{dark state of ri})
is illustrated for the pH$_2$ number density $1\times 10^{21}$cm$^{-3}$
in Fig(\ref{power dependence of PSR}) 
$\sim$ Fig(\ref{r3 profile 2}).
We exhibit dependence of the symmetric output pulse on
the trigger power in the range of   $10^{-12} \sim 1$Wmm$^{-2}$
for $n=1\times 10^{21}$cm$^{-3}$ in Fig(\ref{power dependence of PSR}),
which demonstrates two important features of explosive PSR with
the presence of a large initial coherence:
(1) the highest peak of PSR output is almost independent of the trigger power,
suggesting a sudden, macroscopic release of energy 
(its density $\approx \epsilon_{eg}n$) stored between two levels, $|e\rangle $
and $|g \rangle $,
(2) the onset time of explosive events, which may be called the delay time,
depends on the input trigger
power very weakly, and a linear logarithmic dependence has been confirmed
up to 1 pW mm$^{-2}$ (instantaneous enhancement factor 
$\sim 8\times 10^{21}$ in this case).
A similar logarithmic power dependence of the delay time
has been observed in numerical simulations of 
the single photon superradiance when the system
is subjected to the trigger.

The integrated flux is $\sim |E_{{\rm max}}|^2 \Delta t$ with
$\Delta t$ the time width of explosive event.
$|E_{{\rm max}}|^2=O[\epsilon_{eg}n]$ and
this integrated flux is estimated as
$O[1/\mu_{ge}]$, a quantity independent of
the target number density $n$, if the explosive event occurs.
These figures show dramatic effects of initial
coherence of the dark state.
Observation of explosive events requires
a target length $\gg 1/\alpha_m \propto 1/n$.

Detailed time structure of pulses as observed in 
Fig(\ref{power dependence of PSR}) may
differ if one adopts different available experimental parameters, 
but the output release of energy flux of order $\epsilon_{eg} n$
is universal in explosive PSR events.

\begin{figure}[htbp]
\begin{minipage}{20em}
\includegraphics[width=\textwidth]{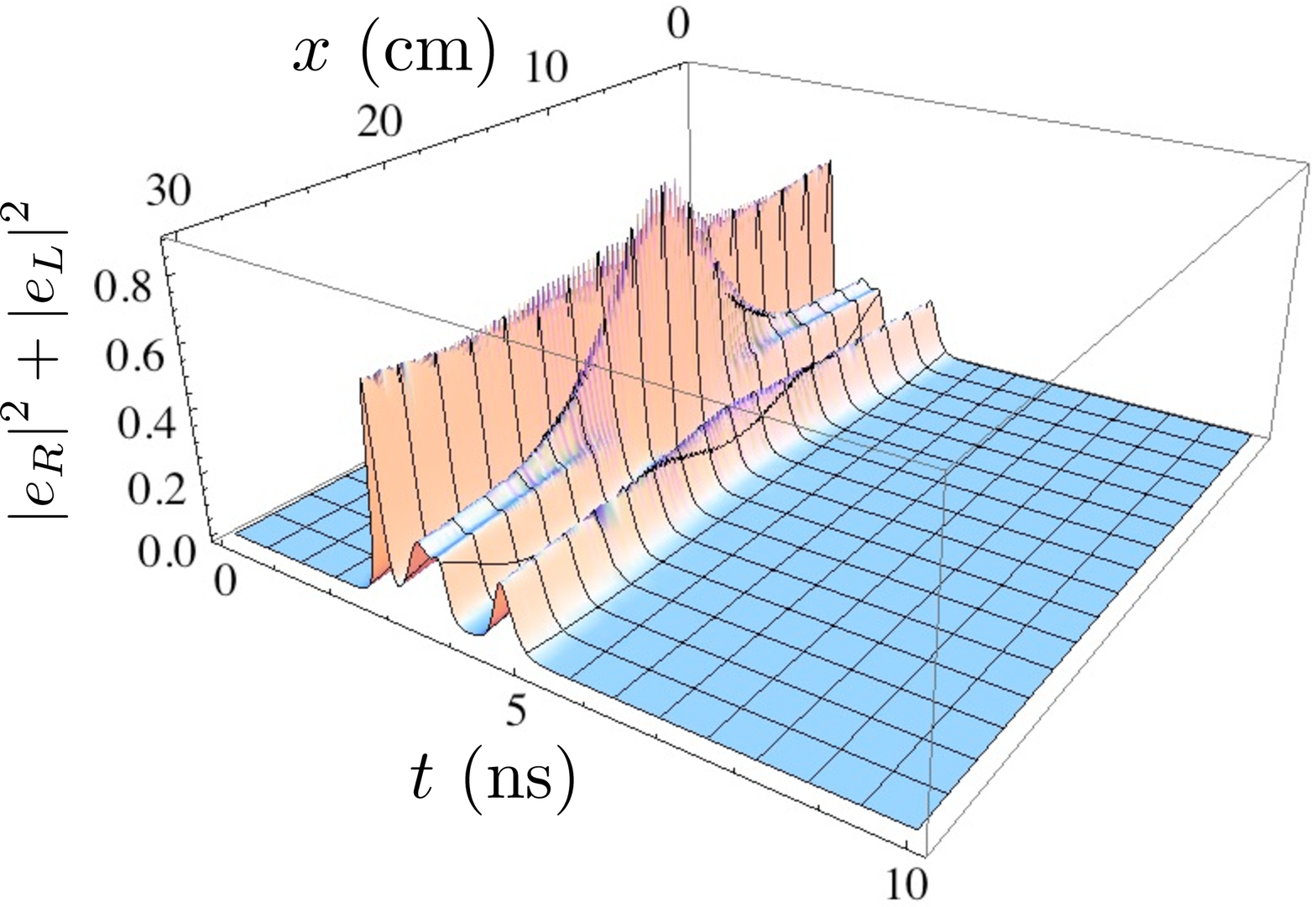}
\caption{\label{energy density profile}
(Color)
 Spacetime profile of dimensionless field energy, $|e_R|^2+|e_L|^2$ for 
 the 1 $\mu$Wmm$^{-2}$ case of Fig(\ref{power dependence of PSR}).}
\end{minipage}
\hspace{2em}
\begin{minipage}{20em}
\includegraphics[width=\textwidth]{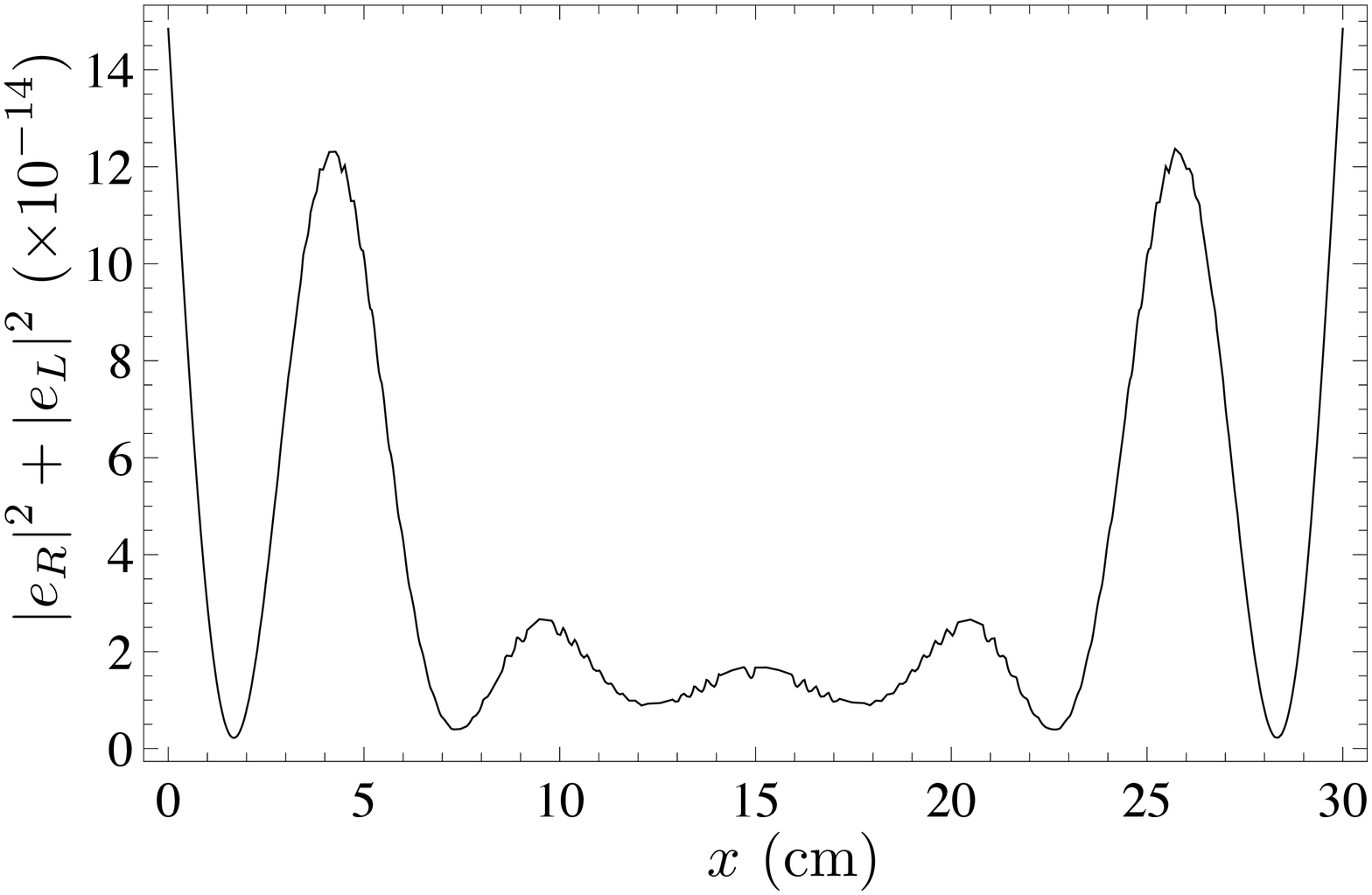}
\caption{\label{energy density profile 2}
 Spatial profile at the latest time, 10 ns after trigger irradiation,
 of Fig(\ref{energy density profile}). Note a large reduction by $O[10^{-13}]$
 in the power scale in this figure.}
\end{minipage}
\end{figure}

Spatial profiles of field fluxes and polarization components,
$r_i$, within the target are illustrated in Fig(\ref{energy density profile}) 
$\sim$ Fig(\ref{r3 profile 2}).
In this parameter set,
about $\sim$ 30 \% of the stored energy $\epsilon_{eg}n$
(the corresponding flux unit being 
$1.2 \times 10^9$ Wmm$^{-2} (n/ 10^{21}{\rm cm}^{-3})$)
still remains in the target much after explosive PSR,
and we observe a seemingly stable target state.
Note that dimensionless fields $|e_i|^2 = |E_i|^2/(\epsilon_{eg}n)$
are plotted in Fig(\ref{energy density profile}) and 
Fig(\ref{energy density profile 2}).

\begin{figure}[htbp]
\begin{minipage}{20em}
\includegraphics[width=\textwidth]{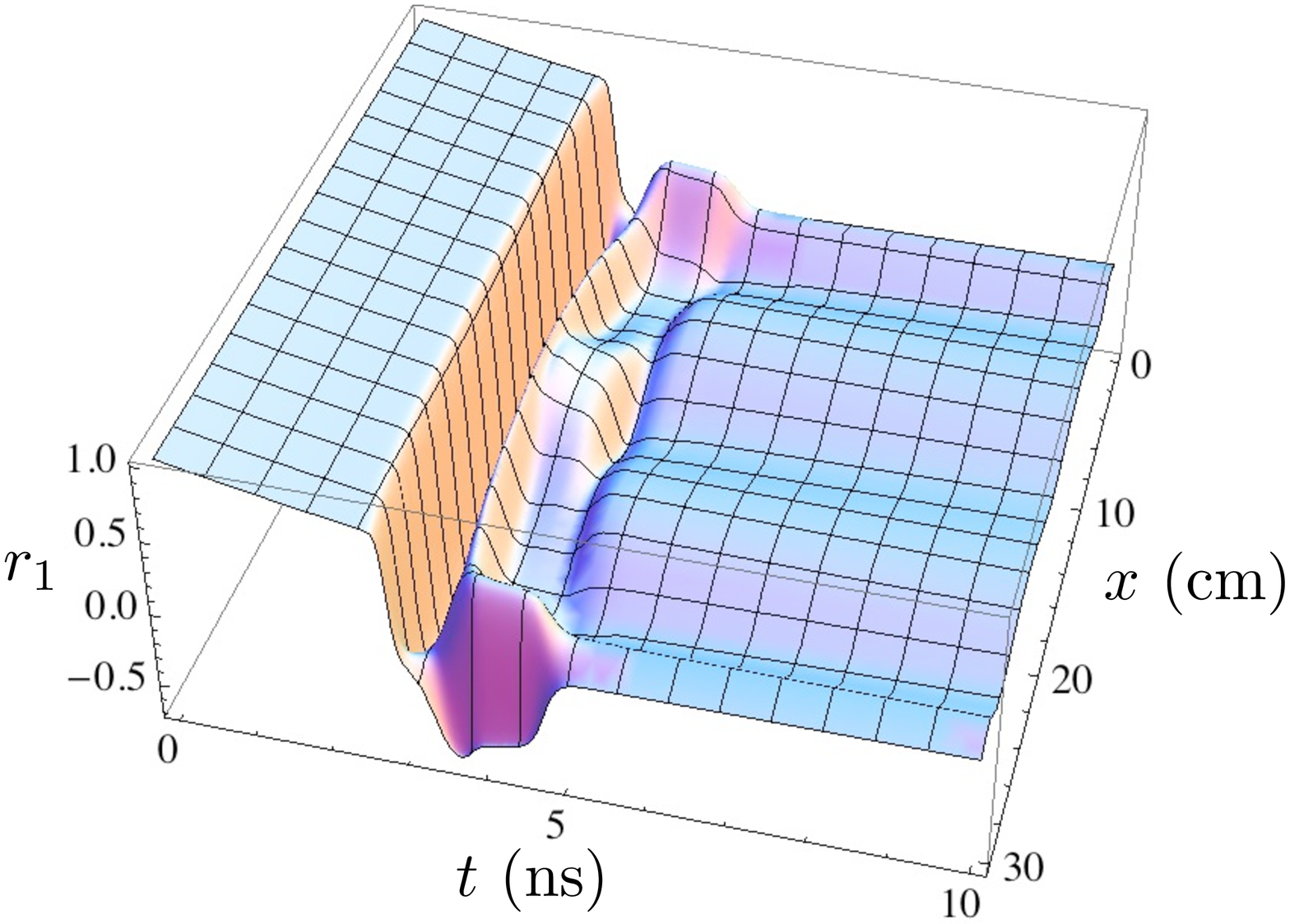}
\caption{\label{r1 profile}
(Color)
 Spacetime profile of $r_1$  for the 1 $\mu$Wmm$^{-2}$ case 
 of Fig(\ref{power dependence of PSR}).}
\end{minipage}
\hspace{2em}
\begin{minipage}{20em}
\includegraphics[width=\textwidth]{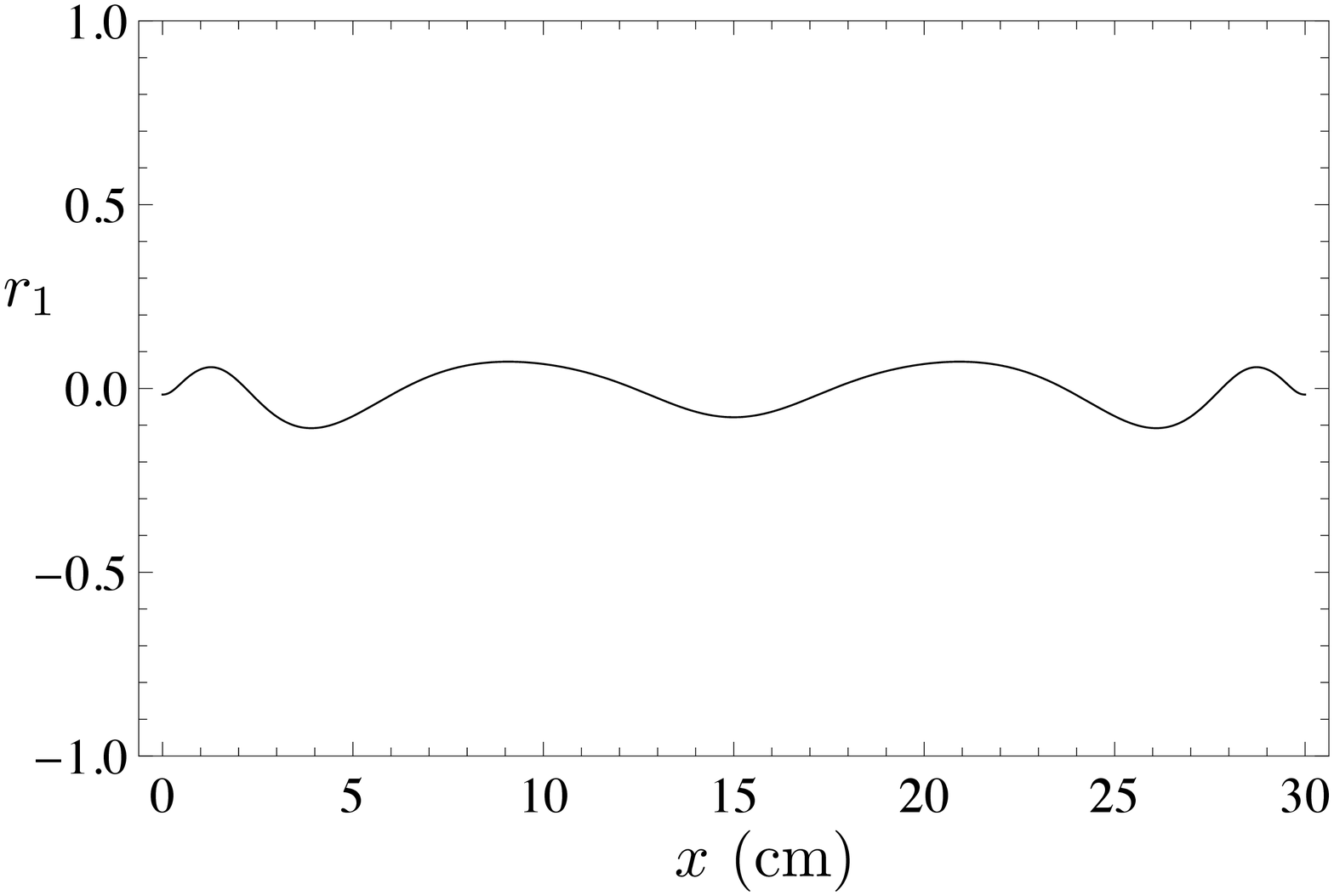}
\caption{\label{r1 profile 2}
 Spatial profile of $r_1$ at the latest time, 10 ns 
 after trigger irradiation, of Fig(\ref{r1 profile}).}
\end{minipage}
\end{figure}

\begin{figure}[htbp]
\begin{minipage}{20em}
\includegraphics[width=\textwidth]{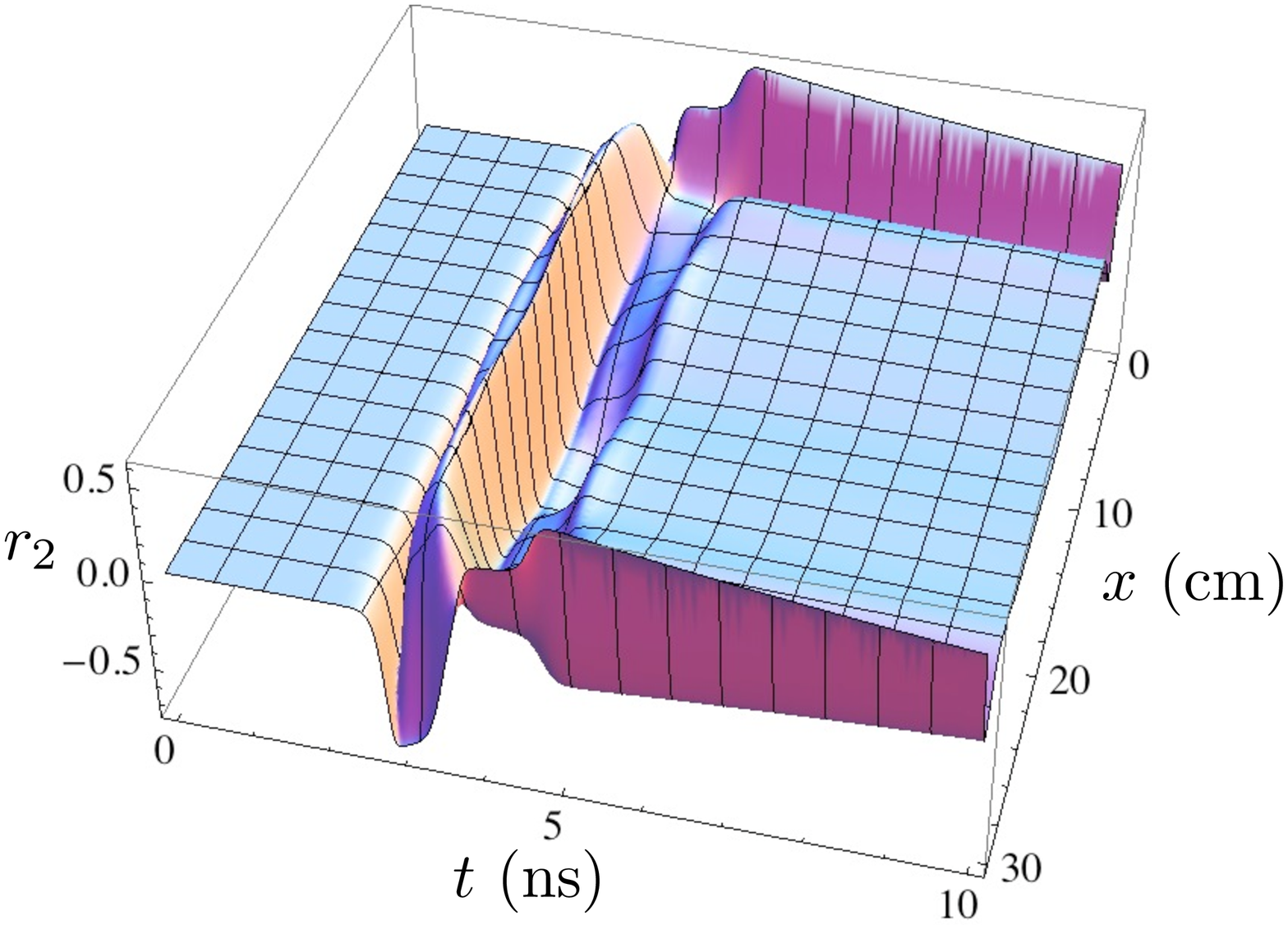}
\caption{\label{r2 profile}
(Color)
 Spacetime profile of $r_2$  for the 1 $\mu$Wmm$^{-2}$ case 
 of Fig(\ref{power dependence of PSR}).}
\end{minipage}
\hspace{2em}
\begin{minipage}{20em}
\includegraphics[width=\textwidth]{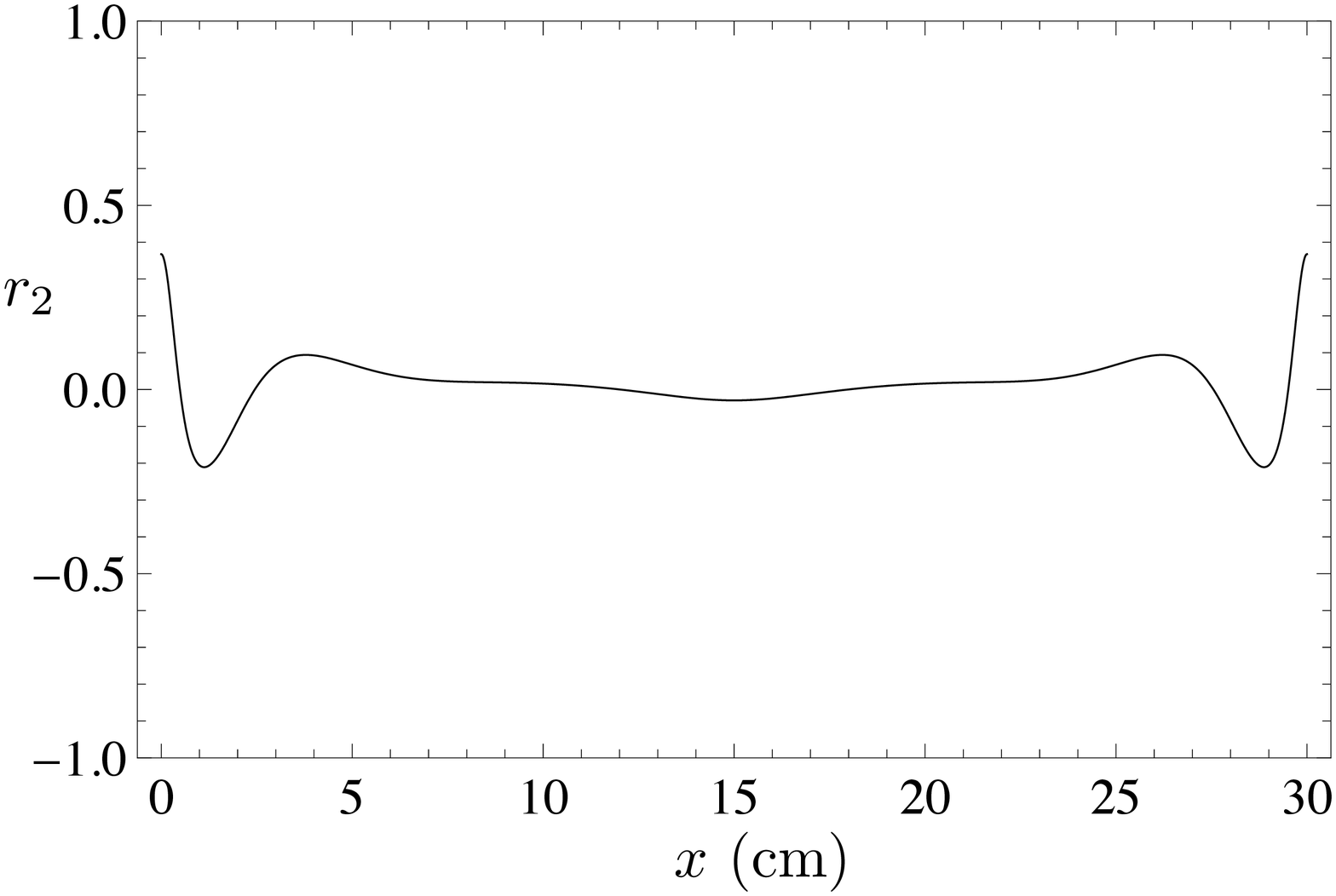}
\caption{\label{r2 profile 2}
 Spatial profile of $r_2$ at the latest time, 10 ns 
 after trigger irradiation, of Fig(\ref{r2 profile}).}
\end{minipage}
\end{figure}

\begin{figure*}[htbp]
\includegraphics[width=30em]{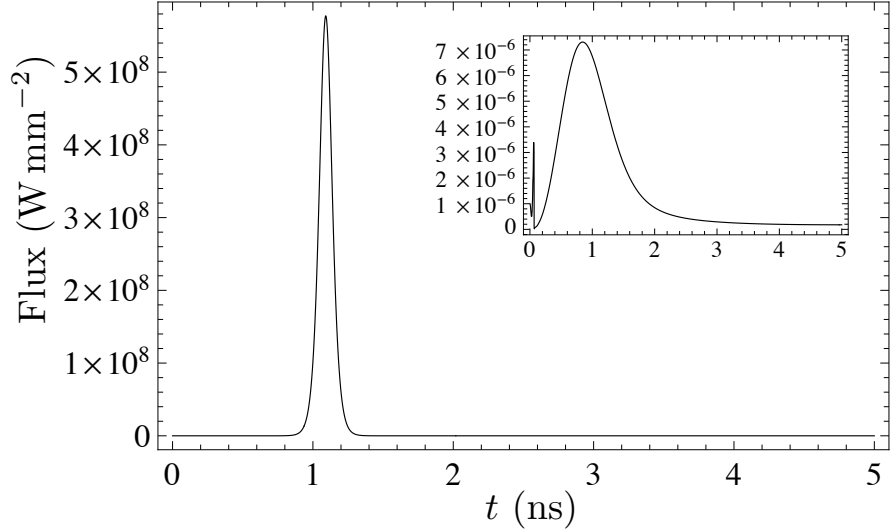}
\caption{\label{large density, small r} 
 Output flux for the solid target number density 
 $2.6 \times 10^{22}\:\mathrm{cm}^{-3}$ of length  2 cm, the trigger power 
 1 $\mu$Wmm$^{-2}$, relaxation times $T_2=10, T_1=10^3$ ns's,
 and smaller population $r_3= - 0.99$ (0.5\% excitation), and 
 $r_3= - 0.996$ (0.2\% excitation) in the inset. The other initial components
 are taken as $r_1 = \sqrt{1-r_3^2}\,,\;r_2=0$.
 Note a large flux scale difference $\sim 10^{14}$ in two plots.}
\end{figure*}

\begin{figure*}[htbp]
\includegraphics[width=30em]{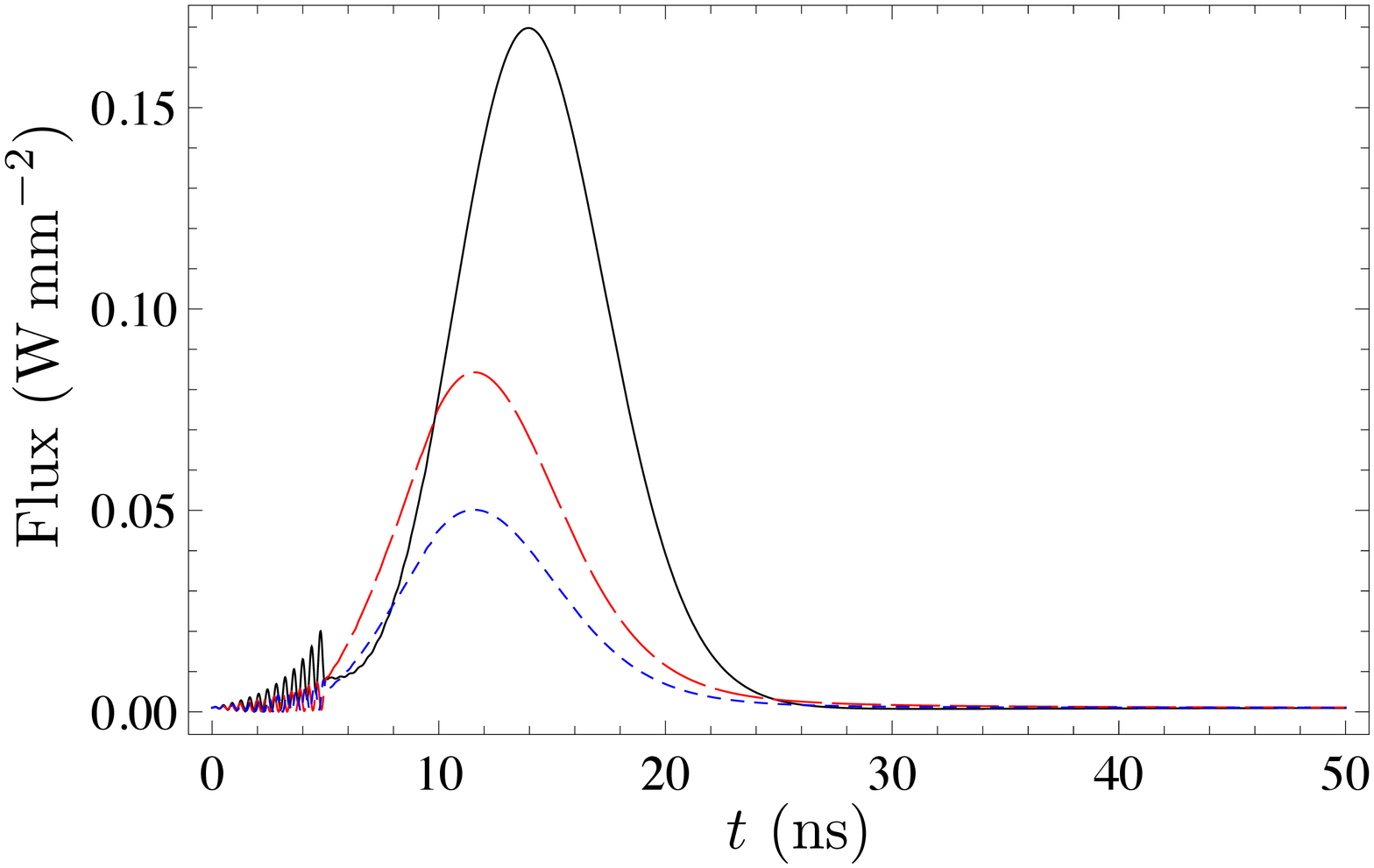}
\caption{\label{output in linear regime}
(Color online)
 Output flux of weak pH$_2$ PSR in the linear regime
 in which the output power $\sim O[10^2] \times$ the trigger power, 
 for initial $(r_3, r_1, r_2) = (0,1,0)$ in solid black, 
 $(1/\sqrt{2}, 1/\sqrt{2},0)$ in dashed red and $(-1/\sqrt{2}, 1/\sqrt{2},0)$
 in dotted blue, using the same set of other parameters:
 $n= 1 \times 10^{20}$cm$^{-3}$, target length = 1.5 m,
 relaxation times $T_2=10, T_1=10^3$ ns's, and the trigger power 1 mWmm$^{-2}$. 
 The output power scales with the trigger power,
 as explicitly checked in the range of 1 $\mu$Wmm$^{-2} \sim 1$ Wmm$^{-2}$.}
\end{figure*}

Result for the solid density of $n=2.6 \times 10^{22}$cm$^{-3}$ and smaller
excitation of  $r_3 \approx -1$ is shown in Fig(\ref{large density, small r}).
There is a threshold of the excitation fraction of $|e\rangle$, located
between 0.2 \% $\sim$ 0.5 \%,
above which dramatic explosive PSR's emerge,
as inferred from comparison of two plots of Fig(\ref{large density, small r}).

So far we mostly showed  explosive outputs in which most of
the stored energy between $|e\rangle $ and $|g\rangle $
is released in a short time $< 10$ns after time delay.
There is however a linear regime under a large initial 
coherence $r_i\,, i=1,2$ in which 
the output flux is amplified in proportion to the trigger power.
For instance, the amplification factor is $\sim 10^2$
in the trigger power range of 1 $\mu$Wmm$^{-2} \sim 1$ Wmm$^{-2}$
for three different choices of initial $r_i$ values
of Fig (\ref{output in linear regime}).
In this figure we show
the output fluxes in the linear regime taking 
as an example the trigger power of 1 mWmm$^{-2}$.
Although not shown in this figure,
the linearity of the output power to the trigger power
has been checked for this set of parameters.

\section{Static remnant and spinorial solitons}
\label{SEC:SOLITON}
In addition to dramatic explosive PSR emission it is also
important to watch remnants after PSR emission,
since previous figures at latest times may be taken to suggest
formation of objects of non-trivial spatial profiles.
Let us derive for this purpose the asymptotic form
of fundamental equations.
We anticipate that both the medium polarization $\vec{r}$
and fields $e_R, e_L$ little change with time in the time region of
$t \gg 1/\alpha_m$ after PSR emission.
By taking vanishing time derivatives, one may eliminate
polarizations $r_i$ in favor of field components and
write profile equations of spatial variation for fields,
\begin{equation}
 e_R' = 2i g e_R +ife_L^*
\,, \hspace{0.5cm}
 e_L' = - 2i g e_L - ife_R^*
\,,
\label{profile eq}
\end{equation}
\begin{eqnarray}
&&
g=g(e_R, e_L) = \gamma_+ -\gamma_-
                \frac{16\gamma_-^2\tau_2^2(|e_R|^2 + |e_L|^2)^2 + 1}
{16\gamma_-^2 \tau_2^2(|e_R|^2 + |e_L|^2)^2+ 64 \tau_1\tau_2|e_Re_L|^2+1 }
\,, 
\\ &&
f=f(e_R, e_L) = 
\frac{4 \tau_2  e_R e_L \left( (4\gamma_-\tau_2(|e_R|^2 + |e_L|^2) - i \right)}
{16\gamma_-^2 \tau_2^2(|e_R|^2 + |e_L|^2)^2+ 64 \tau_1\tau_2|e_Re_L|^2+1 }
\,,
\end{eqnarray}
\begin{equation}
r_3= - \frac{16\gamma_-^2 \tau_2^2(|e_R|^2 + |e_L|^2)^2 + 1}
{16\gamma_-^2 \tau_2^2(|e_R|^2 + |e_L|^2)^2+ 64 \tau_1\tau_2|e_Re_L|^2+1 }
\,,
\label{soliton r3}
\end{equation}
where $'$ indicates the spatial derivative $\partial_{\xi}$.

Despite of complicated field dependent coefficient functions that appear in 
$f,g$,
the structure of profile equation (\ref{profile eq}) is rather simple.
Oscillatory behavior governed by terms $\propto g$
can be eliminated by taking three bilinear forms of 
fields, $|e_R|^2\,, |e_L|^2\,, e_Re_L$:
\begin{equation}
(|e_R|^2 + |e_L|^2)' = 0\,,\  
(|e_R|^2 - |e_L|^2)' = - 4 \Im(f e_R^*e_L^*)\,, \
(e_Re_L)' = -if(|e_R|^2 - |e_L|^2)\,,
\label{flux relation}
\end{equation}
where the function $f$ depends effectively on $e_Re_L$ alone
since the total flux is a constant of integration due to
the first equation of (\ref{flux relation}), hence with
a real constant $e_0$,
$|e_R(\xi)|^2 + |e_L(\xi)|^2 = e_0^2$.
The set of profile equations, (\ref{flux relation}), is transformed into
two equations of phase functions, $\varphi(\xi)\,, S(\xi)$, defined by
\begin{eqnarray}
&&
e_R(\xi)=e_0 \cos \varphi(\xi)
\,, \hspace{0.5cm}
e_L(\xi)=e_0 e^{iS(\xi)} \sin \varphi(\xi)
\,,
\label{def of two phases}
\\ &&
\varphi' =  \frac{2e_0^2\tau_2}
{1+ 16 \gamma_-^2 e_0^4 \tau_2^2 + 16 e_0^4\tau_1 \tau_2 \sin^2(2\varphi) }
\sin (2\varphi)
\label{phase eq 1}
\,,
\\ &&
S' =  \frac{16 \gamma_- e_0^2\tau_2^2}
{1+ 16 \gamma_-^2 e_0^4 \tau_2^2 + 16 e_0^4\tau_1 \tau_2 \sin^2(2\varphi) }
\cos (2\varphi)
\,,
\label{phase eq 2}
\end{eqnarray}
with $\varphi(l/2) = \pi/4\,, S(l/2) = 0\,,
l = \alpha_m L$.
Since $e_i$'s contain four real functions,
the resulting two equations here reflects
a non-trivial self-consistency of the ansatz (\ref{def of two phases}).
A similar equation with $R \leftrightarrow L$ interchanged may
be set up, suggesting another kind of solitons.

Equation for the angle function $\varphi(\xi)$  (\ref{phase eq 1})
is self-contained, and has the following analytic solution under
the boundary condition $e_R(l/2) = e_0/\sqrt{2}$:
\begin{eqnarray}
&&
2e_R^2 - e_0^2 + \frac{16\gamma_-^2 \tau_2^2 e_0^4 +1}{32\tau_1\tau_2 e_0^4}
\ln \frac{e_R^2}{e_0^2 - e_R^2} = - \frac{\xi-l/2}{4\tau_1}
\,.
\end{eqnarray}
Field may decrease exponentially in the central region, like
$e_R^2 \propto \exp[-8\tau_2 e_0^4|\xi-l/2|/(16 \gamma_-^2 \tau_2^2 e_0^4 +1)]$.
One may define
the soliton size by the e-folding factor as
$\xi_s = (16\gamma_-^2 \tau_2^2 e_0^4 +1)/(8\tau_2 e_0^4)$.
The actual soliton size is $x_s=\xi_s/\alpha_m$.

\begin{figure*}[htbp]
\includegraphics[width=30em]{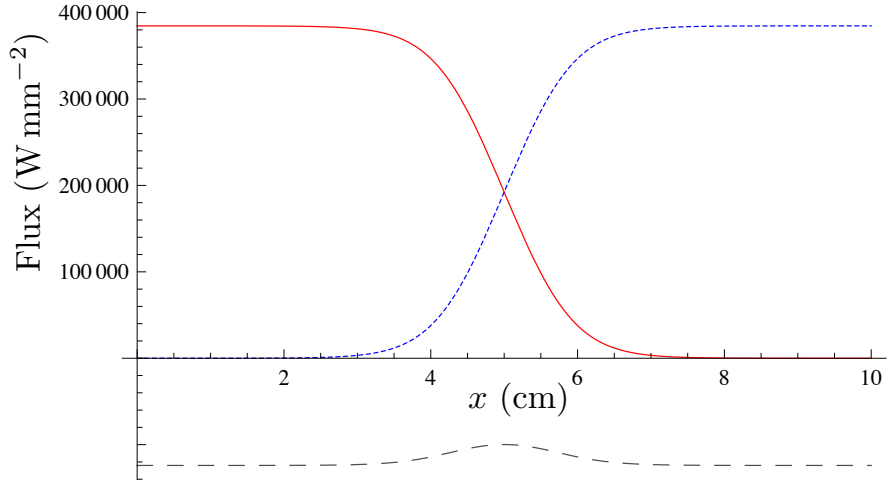}
\caption{\label{soliton structure}
(Color online)
 Profile of fields and $r_3$ of helical absorber soliton.
 $|E_R|^2$ (in red), $ |E_L|^2 $ (in dotted blue), both in the unit Wmm$^{-2}$,
 and $r_3$ in arbitrary unit (in dashed black) are plotted for a case of 
 $n=2.6 \times 10^{22}{\rm cm}^{-3}\,, T_2=20 {\rm ns}\,, T_1=10^3 {\rm ns}$.
 $ r_3 \approx -1$ near edges and $r_3 \approx -0.8$ in the middle.}
\end{figure*}

The spatial variation of $e_R \propto \cos \varphi(\xi)$ is monotonic,
decreasing  or increasing depending on the $\varphi$ region 
of either $[0, \pi/2]$ or $[\pi/2, \pi]$ 
(defined modulo $\pi$).
These two fundamental regions are
separated since $\varphi'=0$ at edges of these regions
due to $\sin (2\varphi) = 0$ there.
One may identify these two solutions as different objects.
Either of fields $e_R\,, e_L$ vanishes at edges
of fundamental regions, but not both.
Solution defined by the fundamental region $[0, \pi/2]$ corresponds to
absorber soliton in which both R- and L-fluxes are absorbed at edges, but
not emitted at the other edges, as illustrated in Fig(\ref{soliton structure})
and Fig(\ref{helical structure}).
This object may be called absorber soliton.
The other fundamental region $[ \pi/2, \pi]$ gives
emitter soliton  which may be realized when the excited
$|e\rangle$ state is sufficiently occupied.
The existence of two types of soliton
condensates is an important result indicating existence of a new
kind of topological soliton whose topology is discussed in Appendix C.

When the target size $L$ is large and $L \gg \xi_s/\alpha_m$,
one may expect a copious production of absorbers and emitters
within the target.
When the target size is smaller than $\xi_s/\alpha_m$,
the target edge effect becomes important
(in general destroying, or blocking its formation of, soliton), and
it may be difficult to create a soliton.

Soliton solution obtained by direct numerical integration of 
(\ref{phase eq 1}) is illustrated in Fig(\ref{soliton structure}) 
along with distribution of the population difference $r_3$.
Solitons are characterized by two end points of $r_3 \approx -1$
and an intermediate region of $r_3 \approx 0$.
It is important to have a long enough target in order to
accommodate many solitons  within the target.
Soliton size can be made smaller
if one can use a larger target number density close to
the solid density.

\section{Conclusion}
In summary,
we derived and numerically  solved the
master equation for time evolution of PSR emission
and formation of field condensates in long dense targets.
We have demonstrated (1) numerical identification of two different types
of PSR events, explosive and weak ones, and
(2) theoretical existence of spinorial solitons stable 
against PSR emission.
Realistic experiments can be designed using
numerical solutions of our master equation.

\vspace{0.5cm}
Note added.

Recently, we became aware of a related work \cite{kalinlin} where
the time evolution of triggered two-photon coherence
is examined.
The authors of \cite{kalinlin} treat the field differently from
the one of our semi-classical approach, which results in
our coherence development time of order several
nano seconds in dense targets, much shorter than
their value.

\begin{acknowledgments}
This research was partially supported by Grant-in-Aid for Scientific
Research on Innovative Areas "Extreme quantum world opened up by atoms"
(21104002)
from the Ministry of Education, Culture, Sports, Science, and Technology.
\end{acknowledgments}

\appendix
\section{Two level effective model interacting with multi-mode fields}
We extend results of Appendix A in
\cite{my-10-10} to the case of multi-mode fields such that
two color problem including propagation effect is properly treated.
This is the most general case of two photon problem.
Its notation in this reference is slightly changed.

\vspace{0.5cm}
{\bf Atomic system}

The state vector of an atom can be expanded in terms of the wave function,
\begin{eqnarray}
&&
| \psi(t) \rangle = c_g(t) e^{-i \epsilon_g t}|g \rangle 
+ c_e(t) e^{-i \epsilon_e t}|e \rangle +
\sum_p c_p(t) e^{-i \epsilon_p t}|p \rangle
\,.
\end{eqnarray}
$ c_{a}(t)$'s are probability amplitudes in an interaction picture
where $\epsilon_a$'s are energies of atomic states.

The atomic system may interact with light fields.
The electric field $E(x,t)$ that appears in 
the hamiltonian via E1 or M1 transition
is assumed to have one vector component alone, namely 
we ignore effects of field polarization.
This is a valid approach under a number of circumstances.
One then decomposes the real field variable $E(x,t)$ into 
Fourier series, $e^{-i\omega_j t}$ times a complex envelope amplitude 
$E_j(x,t)$,
and its conjugate, where $E_j(x,t) $ 
is assumed slowly varying in time,
\begin{eqnarray}
&&
E(x,t) = \sum_j \left( E_j^*(x,t)e^{i\omega_j t} + E_j(x,t)e^{-i\omega_j t}
\right)
\,.
\end{eqnarray}
Each discrete mode $j$ is taken independent.
The most interesting  are the cases of two modes
with $\omega_1 + \omega_2 = \epsilon_{eg}$
and the single mode with  $\omega = \epsilon_{eg}/2$.

The Schr\"{o}dinger equation for a single atom,
\begin{eqnarray}
&&
i \frac{\partial }{\partial  t}|\psi(t) \rangle = (H_0 +
d E )|\psi(t) \rangle
\,,
\end{eqnarray}
with $H_0 $ the atomic hamiltonian,
is used to derive the upper level amplitude $c_p(t)$.
Using
\begin{eqnarray}
&&
i \frac{\partial }{\partial  t} \langle p  |\psi(t) \rangle
= \langle p |(H_0 + d E )|\psi(t) \rangle
\,,
\end{eqnarray}
one has
\begin{eqnarray}
&&
i \frac{dc_p}{dt}e^{- i\epsilon_p t}
=
(d_{pe} c_e e^{- i\epsilon_e t}+d_{pg} c_g e^{- i\epsilon_g t}) E
\,,
\end{eqnarray}
where $d_{ab}$ are dipole matrix elements.
This can formally be integrated to
\begin{eqnarray}
&&\!
\hspace*{-1cm}
c_p(t) = -i\! \int_0^t\! dt' \left(
d_{pe} c_e(t') e^{i \epsilon_{pe}t'}
+ d_{pg} c_g(t') e^{i \epsilon_{pg}t'}
\right) E(x,t')
\nonumber \\ &&\!
= - i\! \int_0^t\! dt' \sum_j \left( d_{pe} c_e(t') 
e^{i \epsilon_{pe}t'} + d_{pg} c_g(t') e^{i \epsilon_{pg}t'}
\right)
\left( E_j^*(x,t')e^{i\omega_j t'} + E_j(x,t')e^{-i\omega_j t'}\right),
\label{upper level coeff}
\end{eqnarray}
with $\epsilon_{ab}=\epsilon_{a} - \epsilon_{b}$ the atomic level energy 
difference. The initial condition $c_p(0) = 0$ is assumed here.

\vspace{0.5cm}
{\bf Markovian approximation and effective two level model} 

The basic strategy of deriving equations for the lower two level amplitudes 
$c_e, c_g$
in a closed form is 
to eliminate atomic variables $c_p$'s related to the upper levels.
This is essentially done by neglecting a long-time
memory effect (the Markovian approximation) and making slowly varying envelope
approximation (SVEA).
The idea of the Markovian approximation is to replace dynamical variables, 
$c_e(t'), c_g(t'), E_j(x,t')$ in the integrand
of eq.(\ref{upper level coeff}), by their 
values at time $t$, neglecting all the past memory effects.
This gives 
\begin{eqnarray}
&&
c_p(t) \approx \sum_j d_{pe}c_e \left( 
\frac{1 - e^{i(\omega_j + \epsilon_{pe})t}}{\omega_j + \epsilon_{pe}}E_j^* 
-  \frac{1 - e^{- i(\omega_j - \epsilon_{pe})t}}
        {\omega_j - \epsilon_{pe}}E_j\right)
\nonumber \\ &&
+ d_{pg}c_g \left( \frac{1 - e^{i(\omega_j + \epsilon_{pg})t}}{\omega_j 
+ \epsilon_{pg}}E_j^*
-  \frac{1 - e^{- i(\omega_j - \epsilon_{pg})t}}
        {\omega_j - \epsilon_{pg}}E_j\right)
\,,
\label{upper level amp}
\end{eqnarray}
which is inserted into
equations for the lower levels 
\begin{eqnarray}
&&
\frac{dc_e}{dt}
= - i \sum_p d_{ep} E(x,t)c_p(t) e^{-i \epsilon_{pe}t}
\,,
\\ &&
\frac{dc_g}{dt}
= - i \sum_p d_{gp} E(x,t)c_p(t) e^{-i \epsilon_{pg} t}
\,.
\end{eqnarray}
Note that $d_{ab} = d_{ba}$ are real by
an appropriate choice of phases.

We ignore rapidly oscillating terms,
keeping in mind the two most important cases of the
mode choice. The result is
\begin{equation}
\frac{d}{dt} \left(
\begin{array}{c}
c_{e}  \\
c_{g}  
\end{array}
\right)= - i{\cal H}_I\left(
\begin{array}{c}
c_{e}  \\
c_{g}  
\end{array}
\right)
\,,
\end{equation}
\begin{eqnarray}
- {\cal H}_I &=&\sum_{jj'}\left(
\begin{array}{cc}
 \mu_{ee}(\omega_j, \omega_{j'}) e^{i(\omega_{j} - \omega_{j'})t} E_j^*E_{j'}
& e^{-i (\omega_{j} + \omega_{j'} - \epsilon_{eg})t} 
\mu_{eg}(\omega_j, \omega_{j'}) E_{j}E_{j'}  \\
e^{i (\omega_{j} + \omega_{j'} - \epsilon_{eg})t} 
\mu_{ge}(\omega_j, \omega_{j'})E_j^* E_{j'}^* 
&   \mu_{gg}(\omega_j, \omega_{j'}) e^{i(\omega_{j} - \omega_{j'}t)}E_j^*E_{j'}
\end{array}
\right)\label{2 level hamiltonian}\\
&\equiv& {\cal E}_j {\cal M}_{jj'}{\cal E}_{j'}\,,\nonumber
\end{eqnarray}
\begin{equation}
 \mu_{ee}(\omega_j, \omega_{j'}) = 
 \sum_p \frac{d_{pe}^2 (2\epsilon_{pe} + \omega_{j} - \omega_{j'})}
{(\epsilon_{pe} + \omega_{j})(\epsilon_{pe} - \omega_{j'})}
\,,\ 
 \mu_{gg}(\omega_j, \omega_{j'}) = 
 \sum_p \frac{d_{pg}^2 (2\epsilon_{pg} + \omega_{j} - \omega_{j'})}
{(\epsilon_{pg} + \omega_{j})(\epsilon_{pg} - \omega_{j'})} 
 \,,
\end{equation}
\begin{equation}
 \mu_{eg}(\omega_j, \omega_{j'}) = 
\sum_p \frac{d_{pe}d_{pg} (\epsilon_{pg} - (\omega_{j}+\omega_{j'})/2)}
{(\epsilon_{pg} - \omega_j)(\epsilon_{pg} - \omega_{j'})}
\,,\ 
 \mu_{ge}(\omega_j, \omega_{j'}) = 
\sum_p \frac{d_{pe}d_{pg} (\epsilon_{pe} + (\omega_{j}+\omega_{j'})/2)}
{(\epsilon_{pe} + \omega_{j})(\epsilon_{pe} + \omega_{j'})}
 \,.
\end{equation}
${\cal E}$ may contain both $E_j$ and $E_j^*$.
We simplify notations below such that
fields are redefined incorporating oscillating factors
$e^{-i\omega_j t}$ in $E_j^+$ etc.

\vspace{0.5cm}
{\bf Single color problem}

We apply the result to the problem of counter-propagating fields $E_R$
and $E_L$ of a single color of $\omega_0 = \epsilon_{eg}/2$.
In the $2\times 2$ hamiltonian (\ref{2 level hamiltonian})
one may use the complex field $E^{+} \equiv E_j e^{-i\omega_j t}$
(the positive energy component corresponding to the photon annihilation
operator) and its
conjugate $E^{-} \equiv E_j^* e^{i\omega_j t}$
(the negative energy component corresponding to the photon creation
operator)
to eliminate phase factors except $e^{\pm i\epsilon_{eg} t}$,
as is done in the main text.
Since each mode is independent,
it separately satisfies the field commutation relation,
necessary for derivation of the quantum field equation,
justifying the result of manuscript for the degenerate case.

More concretely, 
\begin{equation}
- {\cal H}_I =\!\!
\left(\!
\begin{array}{cc}
 \mu_{ee} (E_R^{+}E_R^{-}+E_L^{+}E_L^{-}+E_R^{+}E_L^{-}+E_L^{+}E_R^{-})
& e^{i \epsilon_{eg}t}\mu_{ge}(E_R^{+}E_R^{+}+E_L^{+}E_L^{+}+ 2E_R^{+}E_L^{+})\\
e^{-i \epsilon_{eg} t} 
\mu_{ge} (E_R^{-}E_R^{-}+E_L^{-}E_L^{-}+ 2E_R^{-}E_L^{-})
&   \mu_{gg} (E_R^{+}E_R^{-}+E_L^{+}E_L^{-}+E_R^{+}E_L^{-}+E_L^{+}E_R^{-})
\end{array}
\!\right),
\label{1c two modes hamiltonian}
\end{equation}
\begin{equation}
\mu_{ge} = \frac{2d_{pe}d_{pg}}
{\epsilon_{pg}+\epsilon_{pe}}
\,, \hspace{0.5cm}
\mu_{aa}= \frac{2d_{pa}^2\epsilon_{pa}}{\epsilon_{pa}^2 - \omega_0^2}
\,.
\label{1 mode mu 2}
\end{equation}
(RR) and (LL) terms describe pulse propagation with compression and
splitting, while (RL) terms back-scattering, pair creation, and
pair annihilation. 

\vspace{0.5cm}
{\bf Two color problem}

We may consider for $E_R\,, E_L$  envelopes of two different colors 
of $\omega_i$ with $ \omega_1 + \omega_2 = \epsilon_{eg}$.
Separation of cross mode terms leads to 
\begin{equation}
{\cal H}_I ={\cal H}_{d,R} +{\cal H}_{d,L} + {\cal H}_{12,R}
+ {\cal H}_{12,L}+ {\cal H}_{12,RL}
\,,
\end{equation}
\begin{equation}
-{\cal H}_{d,i} =
\left(
\begin{array}{cc}
\sum_{i,a} \mu_{ee}(\omega_a, \omega_a) E_{i,a}^{+}E_{i,a}^{-}
& e^{i \epsilon_{eg}t} 
\sum_{i,a} \mu_{eg}(\omega_a, \omega_a) E_{i,a}^{+}E_{i,a}^{+}
 \\
e^{-i \epsilon_{eg} t} 
\sum_{i,a} \mu_{ge}(\omega_a, \omega_a) E_{i,a}^{-}E_{i,a}^{-}
&   \sum_{i,a} \mu_{gg}(\omega_a, \omega_a) E_{i,a}^{+}E_{i,a}^{-}
\end{array}
\right)
\,,
\end{equation}
\begin{equation}
-{\cal H}_{12,i} =\!\!
\left(\!\!
\begin{array}{cc}
 \mu_{ee}(\omega_1, \omega_2) E_{i,1}^{+}E_{i,2}^{-}
 + \mu_{ee}(\omega_2, \omega_1) E_{i,2}^{+}E_{i,1}^{-}
& e^{i \epsilon_{eg}t}  
2\mu_{eg}(\omega_1, \omega_2) E_{i,1}^{+}E_{i,2}^{+} \\
e^{-i \epsilon_{eg} t} 
2\mu_{ge}(\omega_1, \omega_2)
E_{i,1}^{-}E_{i,2}^{-}
&   \mu_{gg} (\omega_1, \omega_2) E_{i,1}^{+}E_{i,2}^{-}
 + \mu_{gg}(\omega_2, \omega_1) E_{i,2}^{+}E_{i,1}^{-}
\end{array}
\!\!\right),
\end{equation}
\begin{eqnarray}
-({\cal H}_{12,RL})_{aa} &=&
 \mu_{aa}(\omega_1, \omega_2) E_{R,1}^{+}E_{L,2}^{-} 
+ \mu_{aa}( \omega_2,\omega_1)E_{R,2}^{+}E_{L,1}^{-}\nonumber\\ &&
+ \mu_{aa}(\omega_1, \omega_2)E_{L,1}^{+}E_{R,2}^{-} 
+ \mu_{aa}( \omega_2,\omega_1) E_{L,2}^{+}E_{R,1}^{-}
 \,,
\end{eqnarray}
\begin{equation}
-({\cal H}_{12,RL})_{eg} =
e^{i \epsilon_{eg}t}  
2\mu_{eg}(\omega_1, \omega_2) (E_{R,1}^{+}E_{L,2}^{+} + E_{R,2}^{+}E_{L,1}^{+})
\,,
\end{equation}
\begin{equation}
\mu_{aa}(\omega_1, \omega_2) = \frac{d_{pa}^2(2\epsilon_{pa}
+ \omega_1 - \omega_2)}{(\epsilon_{pa} + \omega_1)(\epsilon_{pa} - \omega_2)}
\,,
\end{equation}
\begin{equation}
\mu_{ge}(\omega, \epsilon_{eg}-\omega) =
\mu_{ge}(\epsilon_{eg}-\omega, \omega) =
\mu_{eg}(\omega, \epsilon_{eg}-\omega) =
\mu_{eg}(\epsilon_{eg}-\omega, \omega) =
\frac{d_{pe}d_{pg}(\epsilon_{pg}+\epsilon_{pe})}
{(\epsilon_{pe}+ \omega)(\epsilon_{pg}-\omega)}
\,.
\label{off-diagonal identity}
\end{equation}

\vspace{0.5cm}
{\bf Bloch equation}

The Bloch vector defined by 
\begin{eqnarray}
&&
\vec{R} = \langle \psi | \vec{\sigma} | \psi \rangle
= {\rm tr}\, \vec{\sigma} \rho
\,, \hspace{0.5cm}
\rho = | \psi \rangle \langle \psi | =
\left(
\begin{array}{cc}
c_e^*c_e &c_e^*c_g  \\
c_g^*c_e & c_g^*c_g
\end{array}
\right)
\,,
\end{eqnarray}
satisfies quantum mechanical equation (disregarding relaxation terms)
\(\:
\partial_t \vec{R} = -i {\rm tr}\; \vec{\sigma}[{\cal H}_I\,, \rho]
\,.
\:\)
Explicit calculation using the Hamiltonian above gives
\begin{eqnarray}
&&
\!\!\!\!\!\partial_t R_1 =
 (\mu_{ee}-\mu_{gg})E^+E^- R_2 
- i\mu_{ge} (e^{i\epsilon_{eg}t}E^+E^+ - e^{-i\epsilon_{eg}t}E^-E^-) R_3 
\,,
\\ &&
\!\!\!\!\!\partial_t R_2 = -(\mu_{ee}-\mu_{gg})E^+E^- R_1
+ \mu_{ge} (e^{i\epsilon_{eg}t}E^+E^+ + e^{-i\epsilon_{eg}t}E^-E^-) R_3 
\,,
\\ &&
\!\!\!\!\!\partial_t R_3 =
 \mu_{ge} \left(
i  (e^{i\epsilon_{eg}t}E^+E^+ - e^{-i\epsilon_{eg}t}E^-E^-  ) R_1
-  (e^{i\epsilon_{eg}t}E^+E^+ + e^{-i\epsilon_{eg}t}E^-E^-  )R_2
\right).
\end{eqnarray}
We suppressed mode index $j$ for simplicity.
Conservation law holds:
\(\:
\partial_t (R_1^2 + R_2^2 + R_3^2 ) = 0
\,.
\:\)

\vspace{0.5cm}
{\bf Field equation}

Commutation relation of fields necessary for
derivation of quantum field equation
\(\:
[E_y(\vec{r}, t)\,, B_z(\vec{r}\,', t)]
= i\partial_x \delta^3(\vec{r} - \vec{r}\,')
\,,
\:\)
is valid for each independent mode.
The double commutator,
\begin{equation}
\partial_t^2 \vec{E}^{\pm} =
- [H \,, [H \,,  \vec{E}^{\pm} \,]\,]
\,, \hspace{0.5cm}
H = \int d^3x ({\cal H}_f + {\rm tr}\, \rho{\cal H}_I) \,,
\end{equation}
\begin{equation}
{\rm tr}\, \rho {\cal H}_I = \langle \psi|{\cal H}|\psi \rangle = 
- (\mu_{ee}|c_e|^2 + \mu_{gg}|c_g|^2)E^+E^-
- \mu_{ge} (c_e^*c_g E^+E^+ + c_g^*c_e E^-E^-)
\,,
\end{equation}
with the field energy density ${\cal H}_f = (\vec{E}^2+\vec{B}^2)/2$,
is calculated as
\begin{equation}
(\partial_t^2 - \vec{\nabla}^2 )\vec{E}_j^{\pm}
=  \vec{\nabla}^2 ({\cal D}_{jj'} \vec{E}_{j'})^{\pm}
\,,
\label{full field eq 2}
\end{equation}
\begin{equation}
-{\cal D}_{jj'}\vec{E}_{j'}^+ =
\left( \frac{(\mu_{ee} + \mu_{gg})_{jj'}}{2}
n + \frac{(\mu_{ee} - \mu_{gg})_{jj'}}{2}
R_3  \right)\vec{E}_{j'}^+ + (\mu_{ge})_{jj'} e^{-i \epsilon_{eg}t}
(R_1-iR_2) \vec{E}_{j'}^-
\,.
\label{rhs of field eq 2}
\end{equation}

\vspace{0.5cm}
{\bf SVEA and dimensionless equations for two color modes}

All terms both in the Bloch and field equations
must have the same oscillatory behavior for
global evolution of polarization and fields.
This gives a phase matching condition of the form
$\omega_1 + \omega_2 = \epsilon_{eg}$ and momentum balance
with $E_R \propto e^{i\omega x}\,, E_L \propto e^{-i\omega x}$.
For time SVEA one may then eliminate the phase factor $e^{\pm i \epsilon_{eg} t}$
in the Bloch equation.
For space SVEA we introduce spatial variation of polarization of the form,
\begin{eqnarray}
&&
R_i = R_i^{(0)} + R_i^{(+)}e^{2i\omega x} + R_i^{(-)}e^{-2i\omega x}
\,.
\label{case of spatial graiting}
\end{eqnarray}

LHS of field equations $\sim -2i \omega (\partial_t \pm \partial_x)E_{R\,, L}$
for the counter-propagating  modes of the same frequency, 
hence (with $\partial_{\pm}\equiv \partial_t \pm \partial_x$)
\begin{eqnarray}
&& 
\partial_+ E_R =
\frac{i\omega}{2}\biggl( (\frac{\mu_{ee} + \mu_{gg}}{2}
n + \frac{\mu_{ee} - \mu_{gg}}{2}R_3^{(0)})E_R 
+ \frac{\mu_{ee} - \mu_{gg}}{2}R_3^{(+)}E_L 
\nonumber \\ &&
\hspace*{1cm}
+ \mu_{ge} 
\left( (R_1-iR_2)^{(0)} E_L^* + (R_1-iR_2)^{(+)}E_R^* \right)
\biggr)
\,,
\\ &&
\partial_- E_L =
\frac{i\omega}{2}\biggl( (\frac{\mu_{ee} + \mu_{gg}}{2}
n + \frac{\mu_{ee} - \mu_{gg}}{2}R_3^{(0)})E_L 
+ \frac{\mu_{ee} - \mu_{gg}}{2}R_3^{(-)}E_R 
\nonumber \\ &&
\hspace*{1cm}
+ \mu_{ge} 
\left( (R_1-iR_2)^{(0)} E_R^* + (R_1-iR_2)^{(-)}E_L^* \right)
\biggr)
\,.
\end{eqnarray}

We introduce the dimensionless unit:
\begin{equation}
(\xi\,, \tau) =  (\alpha_m x\,,\alpha_m t)\,,\
\alpha_m(\omega) = \frac{\epsilon_{eg}}{2}n\mu_{ge}
                   (\omega, \epsilon_{eg}-\omega)\,,\ 
|e_{L, R}^{(1), (2)}|^2 = \frac{|E_{L, R}^{(1), (2)}|^2}{\epsilon_{eg}n}\,,\ 
r_i = \frac{R_i}{n}\,.
\end{equation}
Assume R-mover of frequency $\omega_1$ and L-mover of frequency $\omega_2$
(neither R-mover of frequency $\omega_2$ nor L-mover of frequency $\omega_1$).
Note the universal parameter 
$\mu_{eg}(\omega_1, \omega_2)= \mu_{ge}(\omega_1, \omega_2)$
for any combination of $\omega_1+ \omega_2= \epsilon_{eg}$.
The master equations for medium polarization and fields are
\begin{eqnarray}
\partial_{\tau} r_1^{(0)} &=& 
 4(\gamma_-^{(1)}|e_R|^2  + \gamma_-^{(2)}|e_L|^2)r_2^{(0)}
+ 8 \Im (e_R e_L)r_3^{(0)}
+ 4\gamma_-^{(12)} e_R e_L^* r_2^{(-)} +  4\gamma_-^{(21)} e_L e_R^* r_2^{(+)}
\nonumber \\ &&
- 2i (e_L^2- (e_R^*)^2) r_3^{(+)} - 2i (e_R^2- (e_L^*)^2) r_3^{(-)}
-\frac{r_1^{(0)}}{\tau_2}
\,,
\label{rescaled bloch eq1}
\end{eqnarray}
\begin{equation}
\partial_{\tau} r_1^{(+)} =4\gamma_-^{(12)} e_R e_L^* r_2^{(0)}
- 2i (e_R^2- (e_L^*)^2) r_3^{(0)}
+4(\gamma_-^{(1)}|e_R|^2  + \gamma_-^{(2)}|e_L|^2)r_2^{(+)} 
+  8 \Im (e_R e_L)r_3^{(+)}
-\frac{r_1^{(+)}}{\tau_2}\,,
\end{equation}
\begin{eqnarray}
&&
\partial_{\tau} r_2^{(0)} = 
-4(\gamma_-^{(1)}|e_R|^2  + \gamma_-^{(2)}|e_L|^2)r_1^{(0)} 
+ 8 \Re (e_R e_L)r_3^{(0)}
- 4\gamma_-^{(12)} e_R e_L^* r_1^{(-)}  - 4\gamma_-^{(21)} e_L e_R^* r_1^{(+)}
\nonumber \\ &&
\hspace*{1cm}
+ 2 (e_L^2 + (e_R^*)^2) r_3^{(+)} + 2 (e_R^2 + (e_L^*)^2) r_3^{(-)}
-\frac{r_2^{(0)}}{\tau_2}
\,,
\label{rescaled bloch eq2}
\end{eqnarray}
\begin{equation}
\partial_{\tau} r_2^{(+)} =-4\gamma_-^{(12)} e_R e_L^* r_1^{(0)}
+ 2 (e_R^2 + (e_L^*)^2) r_3^{(0)}
-4(\gamma_-^{(1)}|e_R|^2  + \gamma_-^{(2)}|e_L|^2)r_1^{(+)} 
+  8 \Re (e_R e_L)r_3^{(+)}
-\frac{r_2^{(+)}}{\tau_2}\,,
\end{equation}
\begin{eqnarray}
&&
\partial_{\tau} r_3^{(0)} = 
-8 \left( \Re (e_R e_L)r_2^{(0)}  + \Im (e_R e_L )r_1^{(0)}
\right)
+2i (e_R^2- (e_L^*)^2) r_1^{(-)} +2i (e_L^2- (e_R^*)^2) r_1^{(+)}
\nonumber \\ &&
\hspace*{1cm}
- 2 (e_L^2 + (e_R^*)^2) r_2^{(+)}  - 2 (e_R^2 + (e_L^*)^2) r_2^{(-)}
-\frac{r_3^{(0)}+1}{\tau_1}
\,,
\label{rescaled bloch eq3}
\end{eqnarray}
\begin{equation}
\partial_{\tau} r_3^{(+)} =
2ir_1^{(0)} (e_R^2- (e_L^*)^2) - 2r_2^{(0)} (e_R^2 + (e_L^*)^2)
-8 \left( \Re (e_R e_L)r_2^{(+)}  + \Im (e_R e_L )r_1^{(+)}\right)
-\frac{r_3^{(+)}}{\tau_1}\,,
\end{equation}
\begin{equation}
(\partial_{\tau} + \partial_{\xi})e_R = 
 \frac{ia_1}{2}  (\gamma_+^{(1)}  +  \gamma_-^{(1)} r_3^{(0)} ) e_R
  + \frac{i}{2}\gamma_-^{(12)} r_3^{(+)}e_L
+ \frac{ia_{12}}{2}(r_1^{(0)} - ir_2^{(0)})e_L^*
+ \frac{i}{2}(r_1^{(+)} - ir_2^{(+)})e_R^*
\,, 
\label{rescaled quantum field eq1}
\end{equation}
\begin{equation}
(\partial_{\tau} - \partial_{\xi})e_L = 
 \frac{ia_2}{2} (\gamma_+^{(2)}  +  \gamma_-^{(2)} r_3^{(0)} ) e_L
 + \frac{i}{2}\gamma_-^{(21)} r_3^{(-)}e_R
+ \frac{ia_{21}}{2}(r_1^{(0)} - ir_2^{(0)})e_R^*
+ \frac{i}{2}(r_1^{(-)} - ir_2^{(-)})e_L^*
\,.  
\label{rescaled quantum field eq2}
\end{equation}
\begin{equation}
\gamma_{\pm}^{(a)} = \frac{\mu_{ee}(\omega_a, \omega_a) 
\pm \mu_{gg}(\omega_a, \omega_a)}{2\mu_{ge}}
\,, \hspace{0.5cm}
\gamma_{\pm}^{(ab)} = \frac{\mu_{ee}(\omega_a, \omega_b) 
\pm \mu_{gg}(\omega_a, \omega_b)}{2\mu_{ge}}\,,
\end{equation}
\begin{equation}
a_{i} = \frac{2\omega_i}{\epsilon_{eg}}
\,, \hspace{0.5cm}
a_{ij} = \frac{2\omega_j^2}{\omega_i \epsilon_{eg}}
\,,
\end{equation}
with $\mu_{ab}$ defined by (\ref{1 mode mu 2}).

The single mode equations in the text are readily derived
by taking $\omega_i = \epsilon_{eg}/2, a_i =1, a_{ij}=1$
and all $\gamma_{\pm}^{(ab)} $ $a,b$-independent.

\vspace{0.5cm}
{\bf Pulse compression factor}

We shall estimate pulse propagation effects neglected in the text.
Pulse propagation may be described by ignoring RL mixing terms in the
general master equations.
By taking one mode $e_R$ of one color, the basic propagation equations are
\begin{eqnarray}
&&
\partial_{\tau} r_1 = 4r_3 \Im e_R^2 + 4\gamma_- r_2 |e_R|^2- \frac{r_1}{\tau_2}
\,,
\label{propagation eq1}
\\ &&
\partial_{\tau} r_2 = 4r_3 \Re e_R^2 - 4\gamma_- r_1 |e_R|^2- \frac{r_2}{\tau_2}
\,,
\\ &&
\partial_{\tau} r_3 = -4(r_1 \Im e_R^2 + r_2 \Re e_R^2) - \frac{r_3+1}{\tau_1}
\,,
\\ &&
(\partial_{\tau} + \partial_{\xi})e_R = \frac{i}{2}
\left( (\gamma_+ + \gamma_- r_3)e_R + (r_1-ir_2) e_R^*
\right)
\,.
\label{propagation eqs}
\end{eqnarray}

We shall ignore relaxation terms, taking $\tau_i \rightarrow \infty$.
Results of \cite{my-10-10}  in terms of the area function
follow with the assumption of reality of the function $e_R$. The relation
$r_1 = - \gamma_- r_3$ automatically follows from the consistency of
three Bloch equations.
The fundamental equation of the propagation problem
is given by a single non-linear field
equation in terms of the area function $\theta(\xi, \tau)$:
\begin{eqnarray}
&&
e_R ^2 = \frac{\partial_{\tau}\theta}{4\sqrt{1+\gamma_-^2}}
\,, \hspace{0.5cm}
r_3 =  \pm \frac{\cos \theta}{\sqrt{1+\gamma_-^2}}
\,, \hspace{0.5cm}
r_2 =  \pm \sin \theta
\,,
\\ &&
(\partial_{\tau} + \partial_{\xi})\partial_{\tau}\theta =
\pm \sin \theta \partial_{\tau}\theta
\,.
\end{eqnarray}

Analytic solutions of this non-linear equation  give \cite{my-10-10}

(1) Pulse splitting. The number $N$ of split pulses is given by
the pulse area of the initial flux $F_i(t)$ divided by $2\pi$:
\begin{eqnarray}
&&
N =  
\frac{1}{2\pi}\sqrt{\mu_{ge}^2 + (\mu_{ee} - \mu_{gg})^2/4} 
\int_{-\infty}^{\infty}dy F_i(y) \,.
\end{eqnarray}

(2) Pulse compression.
The pulse of area $< 2\pi$ is compressed by an amount $E$
(result obtained for Lorentzian pulse),
\begin{eqnarray}
&&
E=
\frac{1}{(\alpha_m x \sin (\tilde{\theta}/2) \pm \cos (\tilde{\theta}/2))^2 
+ \sin^2 (\tilde{\theta}/2)}
\,, 
\label{pulse compression}
\\ &&
\tilde{\theta} = \sqrt{\mu_{ge}^2 + (\mu_{ee} - \mu_{gg})^2/4} 
\int_{-\infty}^{t-x}dy F_i(y) \,,
\end{eqnarray}
$\pm$ depending on amplifier (absorber).

We may estimate the pulse compression factor (\ref{pulse compression})
for CW trigger irradiation of duration $t$ in which case
$\tilde{\theta} \sim \beta t$:
\begin{eqnarray}
&&
E = \frac{1}{(\beta t \alpha_m x \pm 1)^2 + (\beta t)^2}
\sim \frac{1}{1 \pm 2\beta t \alpha_m x}
\,.
\end{eqnarray}
In all cases of our interest $\beta t \leq \beta T_2 \ll 1$.
Thus, unless the target length is large enough, 
close to $ 1/(2\beta T_2 \alpha_m)$,
the effect of pulse compression is not large.

\section{Exact and approximate conservation laws}
We focus on the degenerate case of $\omega_1=\omega_2=\epsilon_{eg}/2$.
There are three different classes of exact and
approximate conservation laws:
(1) one exact conservation that holds with finite $T_i$,
(2) one more approximate conservation law that holds in
the $T_1\rightarrow \infty$ limit,
(3) one further approximate conservation law that holds 
in the $T_2 \rightarrow \infty$ limit
($T_1 \gg T_2$ assumed).

The first exact conservation law is 
derived directly from two equations of motion for the field
$e_i$ and it reads as
\begin{eqnarray}
&&
(\partial_{\tau} + \partial_{\xi})|e_R|^2 =
(\partial_{\tau} - \partial_{\xi})|e_L|^2
\label{chirality conservation}
\,.
\end{eqnarray}
An integral form of this conservation for a finite
target of length $L$ ($l=\alpha_m L$ below) is
\begin{eqnarray}
&&
\frac{d}{d\tau}\int_0^{l}d\xi (|e_R|^2 - |e_L|^2) =
- [|e_R|^2 + |e_L|^2]_{\xi=0}^l
\,.
\end{eqnarray}
The integrated quantity of $|e_R|^2 - |e_L|^2$ stored in the target
balances against its flux outgoing from two target ends.
For the symmetric trigger, RHS of this equation vanishes,
and the integral in LHS is a constant of motion.

The second conservation law that holds in the $T_1\rightarrow \infty$
limit is 
\begin{eqnarray}
&&
\partial_{\tau}(r_3 + 4(|e_R|^2 + |e_L|^2)\,) 
+ 4\partial_{\xi}(|e_R|^2 - |e_L|^2) = 0
\,,
\label{energy conservation}
\end{eqnarray}
corresponding to the energy conservation.
The energy density inside the target is a sum of
medium and field energies, $r_3/2 + 2(|e_R|^2 + |e_L|^2)$,
in our dimensionless unit.
The integrated form of this conservation law in the real unit is
\begin{eqnarray}
&&
\frac{d}{dt}\int_0^{L}dx 
\left(\frac{\epsilon_{eg}}{2}R_3 + 2(|E_R|^2 + |E_L|^2 )\right) =
- 2[|E_R|^2 - |E_L|^2]_{x=0}^L
\,.
\end{eqnarray}
The third class of conservation law  
that holds in the $T_2 \rightarrow \infty$ limit is
\begin{eqnarray}
&&
\partial_{\tau} (r_1^2 + r_2^2 + r_3^2) = 0
\,.
\label{matter conservation}
\end{eqnarray}

\section{Helical soliton}
\begin{figure*}[htbp]
\includegraphics[width=30em]{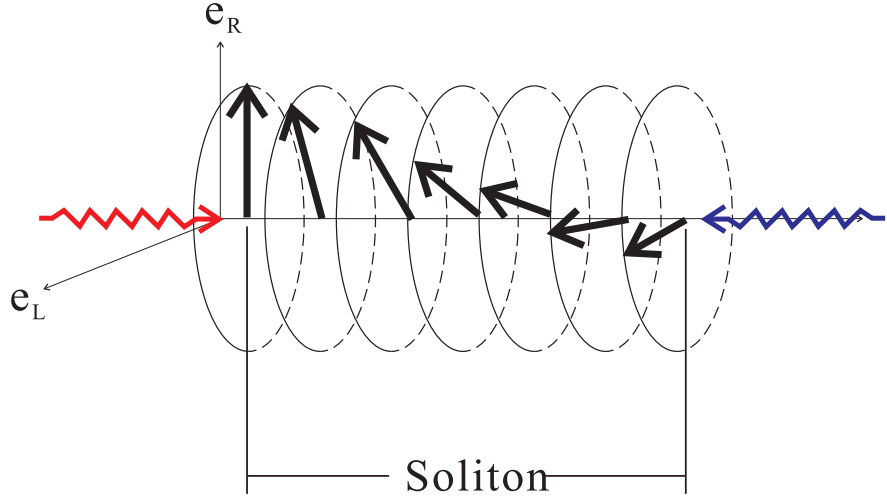}
\caption{\label{helical structure}
 (Color online) Helical structure of absorber soliton. 
 Target region is irradiated from
 both ends by trigger lasers of different colors.
 The $\vec{X}(\xi) = (\cos \varphi, \cos S\sin \varphi, \sin S \sin \varphi)$
 with $\varphi, S$ defined by  eq.(\ref{def of two phases}) may wind.
 In the return trip from the right edge to the left edge,
 not shown here, $\vec{X}(\xi)$ further winds and comes back with 
 $\vec{X}(\pi) = -\vec{X}(0)$ at the left edge, giving a spinor field.
 This is an absorber soliton without emission at two ends.}
\end{figure*}

A new type of topological solitons
may exist, because the basic equation has two
components $\varphi^{(i)}(\xi)\,, i=1,2$, and one can give a topological quantum
number in 1+1 dimensions, as illustrated in Fig(\ref{helical structure}).
For simplicity assume two real component field
$(X,Y)$ and its periodicity with period of
the target length $l (= \alpha_m L)$ or a few times of this length.
We may define the homotopy class \cite{coleman}
of the mapping of a circle $x+iy = l e^{i\xi}\,, 0\leq \xi \leq 2\pi$
in two dimensional real space onto the field space
of the unit magnitude, $X^2 + Y^2 =1$.
The winding number $w$ is defined using the complex
field $Z= X+iY = e^{i\varphi(\xi)}$:
\begin{eqnarray}
&&
w= -i \int_0^{2\pi} \frac{d\xi}{2\pi}Z^*\partial_{\xi}Z
= \frac{\varphi(2\pi) - \varphi(0)}{2\pi}
\,.
\end{eqnarray}
When this winding number is quantized,
$w= n\,, n=0,\pm 1, \pm 2, \cdots$,
the winding number is topologically stable and conserved
during time evolution.

The correspondence to static solutions in Section \ref{SEC:SOLITON}
is as follows.
One considers the real 3-vector field $\vec{X}(\xi)$ of unit length,
$(X,Y,Z) = (\cos \varphi, \cos S\sin \varphi, \sin S \sin \varphi)$
with $\varphi, S$ identified as the phase variables in
static solutions,
and a mapping of unit circle $0\leq \xi \leq 2\pi$ onto $\vec{X}(\xi)$
space. Two solutions of eq.(\ref{phase eq 1}), (\ref{phase eq 2})  
corresponding to two different
fundamental regions, $[0, \pi/2]$ and $[\pi/2, \pi]$,
are connected together at $\xi = \pi/2$.
Then, in the return trip of $\xi = \pi/2 \rightarrow \pi$ 
from the right edge to the left edge
of soliton, the orientation of 
$\vec{X}(\xi)$ is further advanced forward (dictated by continuity
of solutions), and finally comes back with $\vec{X}(\pi) =- \vec{X}(0)$ 
at the left edge.
This means that solutions of $\vec{X}$ are two-valued representation,
namely spinors.


\begin{thebibliography}{99}
\bibitem{another derivation}
The field equation, (\ref{full field eq}) $\sim$ (\ref{rhs of field eq})
and (\ref{full field eq 2}) $\sim$ (\ref{rhs of field eq 2}),
may also be derived by computing macroscopic polarization vector, as is done
in references \cite{narducci} and \cite{my-10-10} for the more restricted case.
\end{thebibliography}
\end{document}